# Moral Hazard on Productivity Among Work-From-Home Workers Amid the COVID-19 Pandemic

Jieun Lee (jieun2@illinois.edu)

Department of Economics,

University of Illinois Urbana-Champaign, USA

**Abstract**

After the outbreak of COVID-19, firms appear to monitor Work-From-Home (WFH) workers more than ever out of anxiety that workers may shirk at home or implement moral hazard at home. Using the Survey of Working Arrangements and Attitudes (SWAA; Barrero et al., 2021), the evidence of WFH workers' *ex-post moral hazard* as well as its specific aspects are examined. The results show that the ex-post moral hazard among the WFH workers is generally found. Interestingly, however, the moral hazard on specific type of productivity, efficiency, is *not* detected for the workers at firms with WFH-friendly policy for long term. Moreover, the advantages & challenges for the WFH culture report that workers with health or disability issues improve their productivity, whereas certain conditions specific to the WFH environment must be met.

# 1. Introduction

The COVID-19 pandemic hit the world in late 2019 and the traditional in-person labor environment was massively threatened as the virus relentlessly spread. Majority of workers and firms worldwide experienced a work-from-home (WFH) situation for the sake of interrupting the dissemination of the virus (Carnevale and Hatak, 2020; Center for Disease Control, 2020; Parker et al., 2020; Savić, 2020; Venkatesh, 2020; Bick et al., 2021; de Lucas Ancillo et al., 2021; Kniffin et al., 2021). Technically, WFH is not a new concept. Before the pandemic, remote workers have worked voluntary, autonomously, and productively at home (Jacobs et al., 1995; Deeprose, 1999; Bloom, 2014; Bloom et al., 2015). Research has found that productivity at a high level was achievable under certain conditions (Neufeld and Fang, 2005; Aboelmaged and Subbaugh, 2012; Bosua et al., 2013). Nonetheless, these findings did not mitigate anxieties for the enforced WFH policy amid the pandemic because they were found in situations where the WFH policy was *not* mandatory. Researchers may have also suffered from a sample selection bias, meaning the results may not be applicable in a general situation.

Because WFH had never been the central culture but was selectively applied to limited cases before the pandemic (McInerney, 1999; Morelli, 1999; Staples, 2001; Mulki et al., 2009; Crawford et al., 2011; Sullivan, 2012; Elshaiekh et al., 2018), little was known about the efficacy of the enforced WFH environment. Responses toward the mandatory WFH trend revealed heterogeneous preferences by position. For example, public sentiment about the new working environment was positive (Dubey and Tripathi, 2020; Bhalla et al., 2021; Zhang et al., 2021), although some remote workers reported suffering from conflicts in work–life balance (Palumbo, 2020). In contrast, managerial sentiment was negative, and managers' perceived risk increased (Ng et al., 2021); global HR leaders' challenges and concerns about workers' commitment and productivity derived from the enforced WFH mandate (Sull et al., 2020); and managers doubted if WFH workers could remain motivated (Parker et al., 2020). Accordingly, Google Search interest in Figure 1 (CNBC, 2020) and the market for the employees monitoring programs[1] have been rapidly increasing since 2020 accompanied with the outbreak of COVID-19 even though they may help workers cope with procrastination (Wang et al., 2021) or lead to legal conflicts

---

[1] Such programs generally provide Firms with "employee activity monitoring, productivity analysis, screen activity recording, browsing history, time tracking" from workers' activities (SoftwareWorld, 2021).

(Ajunwa, 2018). The market is expected to increase even more sharply as shown in Figure 2 (Maximize Market Research, 2021), showing firms' fear on WFH workers' *moral hazard* or *shirking at home* and thus growing demands to supervise them (Brown, 2020; Chemi, 2020; Hatton, 2020; Reports and Data, 2020; Wood, 2020)

While literature on moral hazard before COVID 19 mainly targeted workers in the traditional face-to-face workplace (Butler and Worrall, 1991; Dionne and St-Michael, 1991; Dembe and Boden, 2000; Bolduc et al., 2002; Lewis and Bajari, 2014), studies that summarize surveys on WFH workers' moral hazard have increased since the outbreak of COVID 19 (Foss, 2020; Murray, 2020; Bagley et al., 2021; Irlacher and Koch, 2021; Kawaguchi and Motegi 2021; Oakshott, 2021; Zhuravel and Svenson, 2021; Bisetti et al., 2022), but they lack of models to account for workers' moral hazard. Considering workers' wages are determined by a firm's value (that could be proxied by firm size, firm's financial stability, etc) and the firm's value depends both on the observed efforts and idiosyncratic shocks (such as delays in finding financial investment opportunities due to increased uncertainty in the era of the pandemic), workers' moral hazard could be understood as a standard principal-agent problem and tested by similar frameworks of Chiappori & Salanie (2000) and Lewis & Bajari (2014). Accordingly, this paper finds model-based evidence on WFH workers' moral hazard in the sense that an unexpected hidden factor, other than worker's effort or profile, affects workers' productivities.

The rest of the paper is organized as follows. In Section 2, the model for the source of WFH workers' moral hazard and empirical strategies for analysis are presented. In Section 3 and Section 4, results and discussions are provided, respectively. Finally, I conclude in Section 5.

## 2. Model and Empirical Strategies

*The Source of Moral Hazard*

Consider a firm and a worker and denote Firm and Worker, respectively. Consider a timeline presented in Figure 3. The timeline goes as follows: After the outbreak of COVID 19, both Firm and Worker have spent the transition period on adjusting to the WFH environment (investing in time & equipment to learn how to WFH effectively) and learning about working

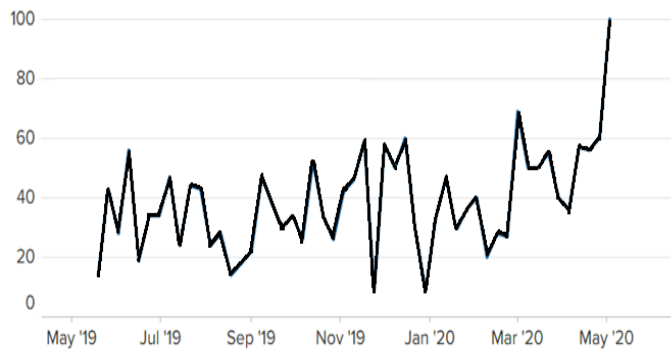

*Figure 1 Google Searches for "Employee Monitoring" Spike*

Source: Google Trends and CNBC (2021)

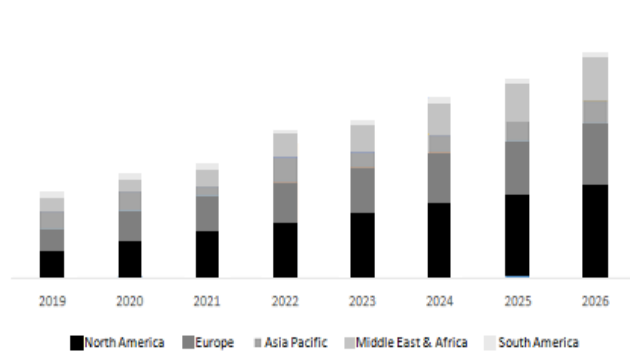

*Figure 2 Global Employee Computer Monitoring Software Market by Region 2019–2026 (USD Million)*

Source: Maximize Market Research (2021)

conditions at home (quality of the IT capital; childcare by the partner or others) and familiarity with virtual communications. Shortly, the situation has settled and WFH policy on a regular basis has begun. The information already learned before the pandemic and learned during the transition period is defined as *ex-ante* information set that contains all information Worker knows pertaining to Firm at the initial period.

*Figure 3 Timeline: Ex-ante & Ex-post*

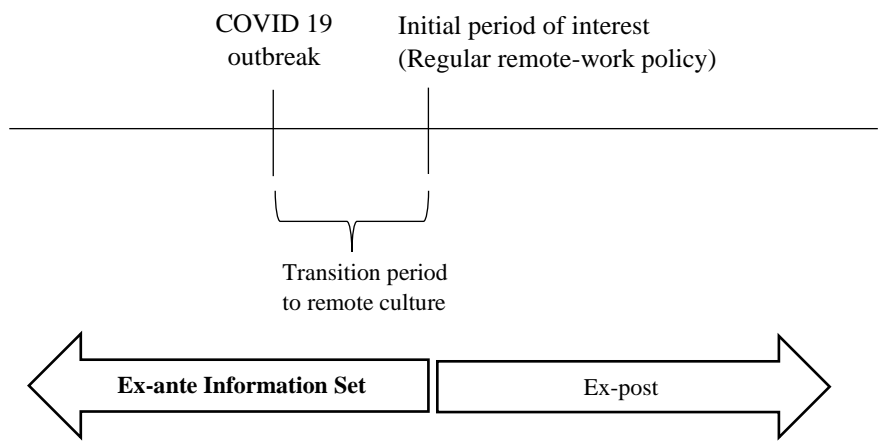

Now similarly to Chiappori & Salanie (2000) and Lewis & Bajari (2014), suppose Firm's value and Worker's productivity equations are decomposed by the ex-ante expectation and random shock, given workers' ex-ante information set $I_f$:

$$\begin{cases} Y_f = E(Y_f | I_f) + \epsilon_f, & (1) \\ P_f = E(P_f | I_f) + v_f, & (2) \end{cases}$$

for all Firms $f = 1, \ldots, F$, where $I_f$ is an ex-ante information set; $Y_f$ denotes Firm's value; $\epsilon_f$ is an unexpected shock on Firm's value; $P_f$ represents Worker's productivity of Worker at Firm $f$; and $v_f$ is a disturbance term for Worker's productivity. Given the observed set of explanatory variables $X_f$, where the conditional expectations of the response variables are assumed to be linear in $X_f$, the equations can be rewritten as

$$\begin{cases} Y_f = X_f \theta_f + \epsilon_f, & (3) \\ P_f = X_f \rho_f + v_f, & (4) \end{cases},$$

where $X_f = (E_{Wf}, C_{Wf}, E_f, C_f)$ with $E_{Wf}$ and $E_f$ being the effort levels of Worker at Firm $f$ and of Firm $f$, respectively, to be effective in a WFH environment. $C_{Wf}$ and $C_f$ include other covariates for Worker at Firm $f$ and for Firm $f$, respectively, such as working conditions at home (quality of the IT capital; childcare by the partner or others), familiarity with virtual communication, level of human capital, health issues, gender, ethnicity, and any other control variables (commute; team collaboration rate at work). Now suppose $(v, \epsilon)$ follows a bivariate normal such that

$$\begin{pmatrix} v_f \\ \epsilon_f \end{pmatrix} \sim N \left( \begin{pmatrix} 0 \\ 0 \end{pmatrix}, \begin{pmatrix} \sigma_v^2 & \sigma_{v\epsilon} \\ \sigma_{\epsilon v} & \sigma_\epsilon^2 \end{pmatrix} \right),$$

where $\sigma_{v\epsilon}$ is supposed to be zero *ex-ante* because they are at different levels of Firm and Worker at the initial period but possibly nonzero *ex-post*. If $\sigma_{v\epsilon} \neq 0$ ex-post, the conditional mean and variance of $v$ conditioning on $\epsilon$ are given as

$$E(v_f | \epsilon_f) = \frac{\sigma_{v\epsilon}}{\sigma_\epsilon^2} \epsilon_f := \delta \epsilon_f, \quad V(v_f | \epsilon_f) = \sigma_v^2 - \frac{\sigma_{v\epsilon}^2}{\sigma_\epsilon^2} := \sigma_\xi^2.$$

Define $\xi_f = v_f - \delta \epsilon_f$ so that $E(\xi_f | \epsilon_f) = 0$ and $V(\xi_f | \epsilon_f) = \sigma_\xi^2$, i.e., $\xi_f$ has a zero mean conditional on $\epsilon_f$. Then equation (3) and (4) can be rewritten as

$$\begin{cases} Y_f = X_f \theta_f + \epsilon_f, & (5) \\ P_f = X_f \rho_f + (\delta \epsilon_f + \xi_f), & (6) \end{cases},$$

where $\delta$ captures the direction and its magnitude of the ex-post moral hazard. If a worker has an incentive to increase the level of productivity ($P_f$) and the estimate for $\delta$ is found to be nonzero and significant, one may interpret it as the *ex-post moral hazard* since the error terms are expected to be independent ex-ante but Worker's productivity is significantly determined by the unexpected hidden factor $\epsilon$, other than Worker's own effort or profiles.

*Prediction for productivity*

One's productivity is found to be multidimensional and thus is inadequate to be approached as a single component. For example, Buglione and Abran (1999) presents an open multi-dimensional model of performance measure by quality factor as well as economic, social, and technical dimensions. Beaton et al. (2009) defines a cost indicator on worker's productivity as a combination of absenteeism and presenteeism and further evaluate productivity using the OMERACT filter. Considering one's performance with a variety of organizational activities, Carlos and Rodrigues (2016) develop a job performance measure particularly in job-related contexts and cultures. (For more examples, see Fraquelli and Vannoni, 2000; Tang and Le, 2007; Kim, 2017; Kim et al., 2018; Soneja and Resham, 2020).

Using a simple average over the variables as the productivity measures does not sound reasonable as it merely gives equal weights over all variables without further consideration on the subtle nature of productivity that depends on the context of the jobs and culture as well as its diverse aspects. Rather, implementing factor analysis and using the predicted values via the most representative factors underlying the productivity structure as the productivity measures is recommended since the prediction by the factor loadings effectively reveals the unique features of each factor. It is expected from the multidimensional nature of productivity that at least two or more factors are found to be significant. A technical summary for factor analysis is provided in the appendix.

*Data Imputation*

Survey data often suffers from a non-negligible number of missing values that may block analysis from utilizing the most information from the original dataset. Moreover, ignoring observations of missing data may yield biased prediction models (Myrtveit et al., 2001). As a remedy, data imputation using predictive models has been introduced and employed in empirical studies (Van Hulse and Khoshgoftaar, 2007; Zhao et al., 2008Sterne et al., 2009; Soley-Bori, 2013; Akande et al., 2017; Jakobsen et al., 2017). Ad hoc remedies to deal with missing values include complete-case analyses that exclude the missing data or mean-value imputation fills the missing values with the mean value from the non-missing values, which may lead to biased estimates (Austin et al., 2021). *Multiple imputation by chained equations (MICE)* is a well-known principled method to resolve this issue (Raghunathan et al., 2001; Van Buuren, 2007; Kyureghian et al., 2011). An iterative series of predictive models contingent upon the other variables in the data fills in the missing values until the algorithm[2] converges (Azur et al., 2011; Wilson, 2021).

A critical assumption for MICE is that the data are missing at random (MAR)[3]. MICE will yield biased estimates if MAR is not met. To test if the data satisfies the missing completely at random (MCAR) condition, Little's test (Little, 1988) is generally used, where the null hypothesis is that the data is MCAR. However, there is no standard statistical test or method to investigate if the data satisfies MAR, but it depends on the contextually appropriate diagnosis (Little and Rubin, 2002).

The standard advice for the number of imputations is two to ten if the number of missing values is not very high (Rubin, 1987; Schafer, 1999; Carpenter and Kenward, 2013), and five is usually set as the default. However, it is subject to a couple of critiques[4]. Moreover, it only addresses the precision and replicability of point estimates (Von Hippel, 2019). Accordingly, a moderate number of imputations is found to be required by a rule of thumb to use the average rate of missingness (Bodner, 2008; White et al., 2011) or a quadratic rule (Von Hippel, 2020). Figure 4 represents the quadratic rule by the fraction of missing information (Von Hippel, 2019).

*Data*

The data contain anonymous micro-level survey data from all waves of the Survey of Working Arrangements and Attitudes (SWAA; Barrero et al., 2021), which spans May 2020 to September 2021. The survey mainly consists of questions for WFH workers in the United States regarding how they feel about the WFH culture and their self-reported efficiency and productivity scales as well as their Firms perspective or policy on WFH.

---

[2] A detailed procedure for MICE algorithm is summarized in Azur et al. (2011).

[3] Rubin (1976) categorized three groups for the missing data mechanism: missing completely at random (MCAR), missing at random (MAR), and missing not at random (MNAR). If the missing data have nothing to do with any values observed or missing in the data, which is often unrealistic, then the data is MCAR. If "any remaining missingness is completely random" (Graham, 2009), but the missing values pertain to the observed data, the data is MAR. If missingness of the data has a significant relationship with its missing values, that is, an underlying system between them determines the missingness, the data is MNAR.

[4] Graham et al. (2007) found that "statistical power for small effect sizes diminished as m became smaller, and the rate of this power falloff was much greater than predicted by changes in relative efficiency". White et al. (2011) "believe that statistical efficiency and power are not enough" and they "want to be confident that a repeated analysis of the same data would produce essentially the same results". Hence "we must consider the Monte Carlo error of our results, defined as their standard deviation across repeated runs of the same imputation procedure with the same data", which clearly decreases as the imputations get large.

*Figure 4 Imputations required by the fraction of missing information*

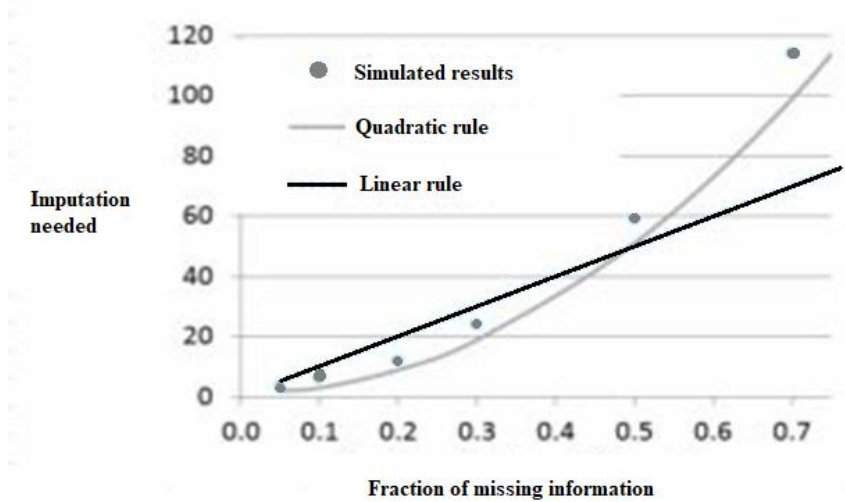

In particular, the variables of interest are (1) WFH-related variables, (2) workers' demographic covariates, and (3) a set of control variables. The WFH-related variables include firms' and workers' respective efforts to adapt to WFH, changes in firms' WFH plans on paid working days for WFH workers after the start of the COVID-19 pandemic, and the current amount of time workers spend in video calls for work, as well as familiarity with WFH culture before the pandemic. In particular, the variables include various kinds of WFH working conditions, including whether workers have their own office, whether workers have support for childcare, device quality, and whether workers have any health issues or disabilities. Moreover, the rate of team collaboration and workers' personal interactions with coworkers or clients are included to see if there are any challenges for workers to keep interactions sustainable. Demographic variables include age, age squared, gender, race, ethnicity, and years of education. The field of industry and commute time that might affect workers' choice of WFH options are grouped as the control variables. Table A1 summarizes the variables codes and their definition in the analysis.

First, I restrict the sample to current WFH workers and further restrict it to wage or salary employees so that workers have representativeness to some extent for each firm. In particular, the data are subsampled by the long-term working environment policy after COVID-19: (1) fully on-site, (2) hybrid (1 to 4 days WFH), or (3) fully remote.

## 3. Results

*Data Imputation*

With the average rate of missing values accounting for 56.17%, diagnosis to acclaim if the data meet MAR condition should be done before implementing MICE. Because there is no standard way to verify it, the cancellation law would be applied: Little's test statistic is 66,444 with the degree of freedom 10,340 and its corresponding p-value 0.000 and thus the data are not MCAR. If the data are MNAR, the distribution for the demographic covariates by the missing and non-missing variable would be quite dissimilar. As Figure A1 in the appendix reports moderately similar patterns by various types of demographic covariates between missing and non-missing datasets, it assures that the data have little chance to be MNAR. Hence the data are thought to be MAR.

Following both the rule of thumb depending on the ratio of missing values (Bodner, 2008; White et al., 2011) and the quadratic rule, the data were imputed by MICE with 60 imputations and 20 iterations for each imputation so that all variables would converge by the iterative algorithm. Table A2 in the appendix provides the summary statistics of the data for analysis.

*Factor Analysis*

Components for worker's productivity include the efficiency or additionally attained by the WFH environment and the promotion possibility in the various scenarios. With these six variables, the possible number of factors ranges from 0 to 3 by the identifiability constraint (Equation A1). Table 1 summarizes the AIC and BIC values (Equations A2, A3) by the number of factors. The smallest value is found in $p = 2$, and thus it is the optimal number of factors subject to the identifiability constraint (Equation A1). Figure 4 visualizes Table 1, where AIC and BIC show a similar pattern. Table 2 shows that both factors are meaningful or representative with their respective sum of squared loadings greater than 1. They explain a moderate extent of the total variance, about 51.4%, with 29.8% and 21.5%, respectively.

Table 1 AIC and BIC

| $p$ | 1 | 2 | 3 |
|-----|-----|-----|-----|
| AIC | -14,367 | **-21,990** | -21,425 |
| BIC | -14,277 | **-21,862** | -21,267 |

Figure 5 AIC and BIC by Number of Factors

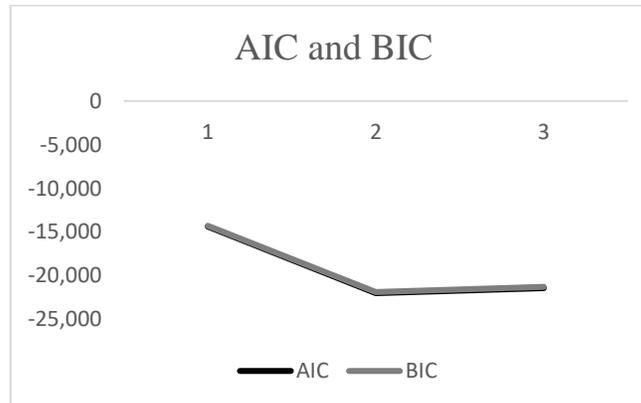

Table 2 Factor analysis results

| | Factor 1 | Factor 2 |
|-----|-----|-----|
| SS loadings | 1.790 | 1.292 |
| Proportion Var | 0.298 | 0.215 |
| Cumulative Var | 0.298 | 0.514 |

Table 3 and its visualization (Figure 5[5]) help label each factor by its loadings on the components for productivity. Factor 1 mainly explains how efficient a worker is in the WFH environment compared with the traditional working condition or relative to the expectation before the pandemic and how much extra efficiency is due to the time saved by not commuting. Hence the most remarkable property of Factor 1 is the WFH efficiency in productivity (*efficiency* for short). Factor 2 mostly explains workers' perceived probabilities of being promoted compared with those of other coworkers who work in person more than the respondent, by one day or by the extremely different pattern. Because promotion is a strong signal representing workers' ability or competence in the workplace (Munjuri, 2011; Claussen et al., 2014; Arokiasamy et al.,

---

[5] Variable codes and their corresponding definitions are provided in Appendix Table A1.

2017), Factor 2 may be interpreted as the WFH competence in productivity (*competence* for short).

*Table 3 Factor analysis results: loadings to each variable*

| Definition | Factor 1 | Factor 2 | Uniqueness |
|---|---|---|---|
| How efficient are you WFH during COVID, relative to on Firm premises before COVID? (%) | 0.922 | 0 | 0.149 |
| Relative to expectations before COVID, how productive are you WFH during COVID? (%) | 0.611 | 0 | 0.622 |
| How much of an increase in your chance of a promotion would working from home one more day per week than your coworkers cause? | 0 | 0.829 | 0.312 |
| How much of an increase in your chance of a promotion would working from home 5+ days a week while your coworkers work on the Firm premises 5+ days a week cause? | 0 | 0.772 | 0.398 |
| How much of your extra efficiency when working from home is due to the time you save by not commuting? | 0.739 | 0 | 0.453 |
| Percent of commute time savings spent working on primary or current job | -0.113 | 0 | 0.984 |

The bivariate kernel density estimations[6] for productivity factors, efficiency, and competence by political preference, years of education, long-term policy on WFH, and industry along with gender and ethnicity support the multidimensionality of productivity with which the factors in productivity are correlated to some extent but with which they are not always concordant. Moreover, the distributions are remarkably heterogeneous based on the variables of gender and ethnicity. Some groups had high efficiency and competence, and others were the opposite; or a polarization of high efficiency and competence and low efficiency and competence within a group was found. This indicates that the distribution of productivities hinges on the variables of interest as well as workers' demographic features, which needs to be interpreted carefully along with the context.

*Regressions*

Table 4 shows the significant and positive effects of worker productivity types on wage, where efficiency shows three times higher values than competence. This reveals workers' ex-post incentive to increase any type of productivity. Table 5 shows the results from the two-stage regressions, where the first stage is for the Firm value equation, and the second stage is for the worker productivity equation.

---

[6] Due to the heavy volume, the figures are available upon request.

*Figure 6 Factor Analysis Results: Loadings*

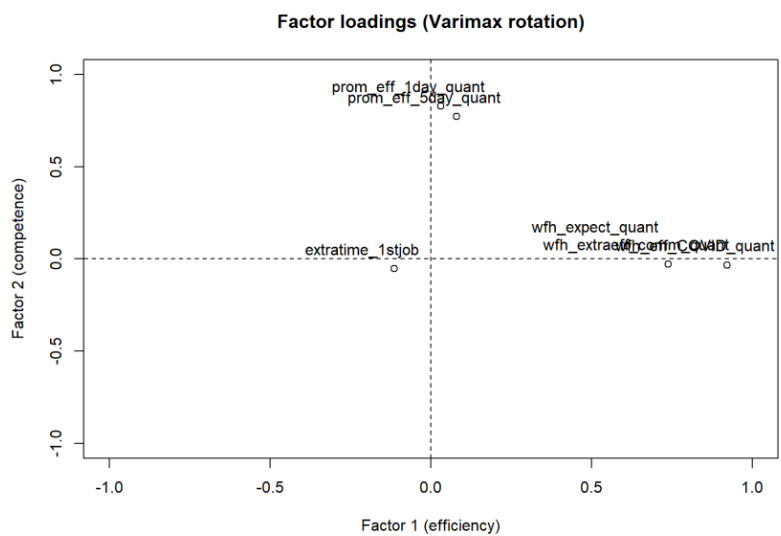

The residuals obtained in the first stage would be another explanatory variable in the second stage and give the estimates for the parameter of interest, the moral hazard coefficient. The first model, denoted (1), represents the model for workers whose Firm has a specific WFH policy for the long term: fully on-site, hybrid (1 to 4 days WFH), or fully remote. The second model, denoted (2), is a restricted model for workers working at a firm with a long-term policy on WFH being fully on-site and labeled as an on-site-favorable firm model. The third model, denoted (3), is another restricted model for workers working at a firm with a long-term policy on WFH being fully remote and labeled as a WFH-friendly firm model.

*Table 4 Wage equation results*

| Variables | Definition | Estimates (Std. Error) |
|---|---|---|
| Intercept | Constant term | 4.565*** |
|  |  | (0.006) |
| Prod.predFactor1 | Predicted score: *Efficiency* in productivity | 0.153*** |
|  |  | (0.006) |
| Prod.predFactor2 | Predicted score: *Competence* in productivity | 0.052*** |
|  |  | (0.007) |
|  | Residual standard error (df) | 0.703 (13,918) |
|  | Multiple R-squared (Adjusted) | 0.044 (0.044) |
|  | F-statistic (df) | 323.2*** (2, 13,918) |

*Regression Results in the First Stage: Firm's Value Equation*

The columns under Stage 1 show the regression results in the first stage for the firm's value equation. Firm's value is proxied by the number of workers in the main team and labeled as firm size. The main results for WFH-related variables are as follows: across the models, workers' efforts to be effective in a WFH environment are significant and positive, confirming that workers' efforts benefit Firm size amid the pandemic. However, the firm's efforts show a significantly negative effect on the firm size across the models, indicating the efforts to adapt to a new environment are a cost to firms.

As predicted, the current video calls percentage are insignificant because it is mainly caused by the shock and works as a default. In contrast, the pre-COVID-19 video call percentage before the pandemic is significantly positive. As such, working culture might have helped firms adapt to the sudden pandemic situation more easily. Firms with workers who have a nice IT capital such as high internet quality, report a greater size for Model 1 and 2. Firms with workers who have their own offices at home tend to have smaller firm sizes. Presumably, some tasks have become unnecessary when transitioning to the WFH environment, which led to a decrease in firm size to some extent. The age of workers' children is also significantly positive. Additionally, team collaboration and worker preference on interpersonal interactions with coworkers or clients are reported to have significant positive effects. Years of education also have a significantly positive effect, as expected.

*Regression Results in the Second Stage: Worker's Productivity Equation*

The regression results in the second stage for the two types of productivity are reported in Table 5. The key finding is that workers' ex-post moral hazard coefficient is found significantly nonzero except for efficiency of workers at WFH-friendly firms. This approves a hidden factor that a random shock on the firms' value does affect workers' productivity ex-post, or workers utilize the shock on the firm to increase their productivities.

Other WFH-related variables also show interesting results. The effort variables of firm and worker to be adjusted to WFH environment have significantly positive effects, implying they help improve productivities. An increase in a firm's planned share of paid working days WFH is significantly positive for Model 2, implying productivity specific to the WFH environment can be learned and improved as firm policies support WFH culture more than before. The frequencies on virtual communications are significant and in general positive, confirming virtual meetings as the representative communication tool for WFH culture. The effects of having an office at home are significant but different depending on the type of productivity. Having an at-home office negatively affects efficiency but positively affects competence. This implies that higher efficiency is subjective to surroundings with coworkers, while competence is enhanced when an independent workplace is provided.

Internet quality is significantly positive on productivities, while the variables on efficiency are twice as high as those on competence, implying efficiency requires more equipment for better information and communication technology capital than competence. WFH culture also shows a great advantage for those who have a health issue or disability, where they show higher efficiency and competence working at home than those who do not. Competence increases as the children get older across the models because time and effort to care for young children may function as a substitute for competence when working at home. Teamwork seems to be a challenge for the WFH environment because it is significantly negative on productivities. At the same time, workers' preference regarding interactions with coworkers or clients has still positive effect on competence. Considering the peculiarity of the WFH environment compared with traditional work premises, this suggests that a team collaboration system specific to the WFH environment may be needed.

Furthermore, demographic features on the productivities report: Gender effect, where the reference group is female, shows that male workers have significantly less efficiency than female workers, whereas they have significantly more competence (visualized in Figure 7). This implies that female and male workers have different adaptive capacities for productivity while working from home. Particularly, male workers at WFH-friendly firms show the least loss in efficiency and the most gain in competence. Race and ethnicity also have effects on efficiency but not on competence (Table A3). The loss in efficiency is mostly found in the workers at on-site favorable firms, while workers at WFH-friendly firms show positive effects across races (visualized in Figure 8). A group of Hispanic (of any race), Native American or Alaska Native, and White (non-Hispanic) shows dissimilarity in effects on efficiency by the firm type: the patterns of the effects of workers at on-site favorable firms and those at WFH-friendly firms are quite dissimilar. This implies that firms with different policy on WFH have very different working environments with respect to efficiency for these races. Meanwhile, Asian and Native Hawaiian or Pacific Islander show similar effects on efficiency within the race, implying the less dissimilarity in working environments of different types of firms.

Years of education are only significantly positive with efficiency. Age is significantly positive only for efficiency, implying efficiency can be learned and the squared age in years shows significant but little diminishing returns to scale. The results for industries, where the reference group is Agriculture, are visualized in Figure 9 and fully reported in Table A3. The patterns for efficiency and competence look quite different, mostly negative effects are found in efficiency while the opposite are presented in competence. For efficiency, some industries such as Arts and entertainment, Finance and insurance, Health care and social assistance, Hospitality and food services, Mining, Professional and firm services, Transportation and warehousing, and Utilities show heterogeneous effects by the type of firm within the same industry. This may indicate the rise of disaggregation for the job tasks within the same industry, depending on the nature of the job whether it needs more traditional working environment or the remote one. For Competence, most of the industries show positive effects on competence. The highest competence is found at the WFH-friendly firms in the Wholesale trade industry and at the on-site favorable firms in the Hospitality and food services, whereas the least competence is found at the WFH-friendly firms in the Mining industry and the on-site favorable firms in the Information industry. This suggests which types of industry fit better for the traditional or remote working environments with respect to competence. For example, the Mining industry may not fit the best with the WFH culture.

## 4. Discussion

In general, the workers' (ex-post) moral hazard induced by the unexpected shock on the firms' value is found. It is remarkable, however, that the moral hazard on efficiency is *not* detected for the workers at firms with WFH-friendly policy for long term. This may indicate that the long-term policy favorable to WFH culture may discourage workers to implement moral hazard on efficiency possibly because workers are motivated to regard home as a regular workplace, whereas workers at on-site-favorable firms may consider WFH temporary and want

*Table 5 Regression results*

| Dependent Variables | Stage 1 | | | Stage 2 | | | | | |
|---|---|---|---|---|---|---|---|---|---|
| | # Employees in the main work team | | | Predicted score: *Efficiency* in productivity | | | Predicted score: *Competence* in productivity | | |
| Variables | (1) All | (2) On-Site Favorable | (3) WFH-friendly | (1) All | (2) On-Site Favorable | (3) WFH-friendly | (1) All | (2) On-Site Favorable | (3) WFH-friendly |
| A constant term | -14.800*** | -15.253*** | -19.460*** | -2.262*** | -2.911*** | -1.089*** | -0.662*** | -1.032*** | -0.629* |
| | (3.007) | (4.973) | (6.433) | (0.157) | (0.254) | (0.349) | (0.157) | (0.257) | (0.342) |
| **First-stage residuals (ex-post moral hazard)** | | NA | | 0.001*** | 0.002** | 0.001 | 0.004*** | 0.004*** | 0.003*** |
| | | | | (0.000) | (0.001) | (0.001) | (0.000) | (0.001) | (0.001) |
| *WFH-related variables* | | | | | | | | | |
| Hours invested in learning how to WFH effectively | 0.064*** | 0.064*** | 0.073*** | 0.002*** | 0.002*** | 0.003** | 0.001*** | 0.002*** | 0.000 |
| | (0.007) | (0.010) | (0.019) | (0.000) | (0.001) | (0.001) | (0.000) | (0.001) | (0.001) |
| Percent of money invested in equipment or infrastructure enabling WFH that was paid for or reimbursed by Firm | -0.054*** | -0.051*** | -0.044*** | 0.004*** | 0.005*** | 0.003** | 0.001*** | 0.000 | 0.000 |
| | (0.004) | (0.007) | (0.011) | (0.000) | (0.000) | (0.001) | (0.000) | (0.000) | (0.001) |
| $\Delta$ Firm planned share of paid working days WFH after COVID (%) | -0.005 | 0.003 | -0.001 | 0.000 | 0.001* | 0.000 | 0.003*** | 0.003*** | 0.002*** |
| | (0.005) | (0.009) | (0.011) | (0.000) | (0.000) | (0.001) | (0.000) | (0.000) | (0.001) |
| Percentage of normal working day spending on video calls | -0.009 | -0.001 | -0.017 | 0.004*** | 0.004*** | 0.004*** | 0.006*** | 0.007*** | 0.004*** |
| | (0.006) | (0.010) | (0.013) | (0.000) | (0.000) | (0.001) | (0.000) | (0.001) | (0.001) |
| Percentage of normal working day spending on video calls (before pandemic) | 0.070*** | 0.077*** | 0.063*** | 0.004*** | 0.003*** | 0.003*** | 0.000 | -0.001* | 0.002** |
| | (0.007) | (0.010) | (0.016) | (0.000) | (0.001) | (0.001) | (0.000) | (0.001) | (0.001) |
| Has their own room (not bedroom) to work in while WFH during COVID)(yes=1) | -1.295*** | -1.012** | -1.174* | -0.093*** | -0.096*** | -0.067* | 0.094*** | 0.079*** | 0.076** |
| | (0.294) | (0.478) | (0.638) | (0.015) | (0.024) | (0.035) | (0.015) | (0.025) | (0.034) |
| Hours of childcare per week provided by partner others (Eg. grandparents, babysitter) | -0.007 | -0.003 | 0.007 | -0.001** | -0.001*** | -0.001 | 0.001*** | 0.001** | 0.001*** |
| | (0.006) | (0.009) | (0.011) | (0.000) | (0.000) | (0.001) | (0.000) | (0.000) | (0.001) |
| Internet quality - Fraction of time that internet works | 3.247** (1.505) | 5.700** (2.775) | 1.527 (2.589) | 1.475*** (0.079) | 2.001*** (0.142) | 1.075*** (0.140) | 0.557*** (0.079) | 0.722*** (0.144) | 0.572*** (0.137) |
| Have a health problem or a disability which prevents work or which limits the kind or amount of work -No (reference: Yes) | 3.822*** (0.375) | 4.795*** (0.548) | 2.163** (0.947) | -0.095*** (0.020) | -0.100*** (0.028) | -0.119** (0.051) | -0.163*** (0.020) | -0.191*** (0.028) | -0.180*** (0.050) |
| Currently live with children under 18 -- categorical by youngest's age | 1.488*** (0.126) | 1.310*** (0.201) | 1.815*** (0.264) | -0.002 (0.007) | -0.002 (0.010) | 0.016 (0.014) | 0.035*** (0.007) | 0.042*** (0.010) | 0.031** (0.014) |
| Percentage of tasks that require collaboration as part of a team | 0.032*** (0.005) | 0.027*** (0.008) | 0.036*** (0.011) | -0.001*** (0.000) | -0.001** (0.000) | -0.001** (0.001) | -0.006*** (0.000) | -0.006*** (0.000) | -0.006*** (0.001) |
| Enjoy personal interactions with coworkers/customers, clients, or patients at Firm's worksite -- ordinal | 0.354*** (0.039) | 0.296*** (0.064) | 0.389*** (0.078) | 0.002 (0.002) | -0.002 (0.003) | 0.000 (0.004) | 0.003 (0.002) | 0.008** (0.003) | 0.013*** (0.004) |
| *Demographic & Control covariates* | | | | | | | | | |
| Male (reference: female) | 2.309*** (0.317) | 1.707*** (0.527) | 2.469*** (0.640) | -0.058*** (0.017) | -0.029 (0.027) | -0.015 (0.035) | 0.054*** (0.017) | 0.060** (0.027) | 0.085** (0.034) |
| Years of education | 0.725*** (0.074) | 0.779*** (0.117) | 0.593*** (0.159) | 0.013*** (0.004) | 0.023*** (0.006) | -0.013 (0.009) | 0.000 (0.004) | 0.004 (0.006) | 0.005 (0.008) |
| Age in years | 0.059 (0.118) | -0.046 (0.195) | 0.211 (0.239) | 0.026*** (0.006) | 0.035*** (0.010) | 0.000 (0.013) | -0.008 (0.006) | -0.008 (0.010) | -0.004 (0.013) |
| Age squared in years | 0.001 (0.001) | 0.002 (0.002) | -0.001 (0.003) | 0.000*** (0.000) | -0.000*** (0.000) | 0.000 (0.000) | 0.000* (0.000) | 0.000 (0.000) | 0.000 (0.000) |
| Commute time (mins) | 0.049*** (0.004) | 0.050*** (0.007) | 0.040*** (0.011) | 0.000 (0.000) | -0.000 (0.000) | 0.000 (0.001) | 0.000 (0.000) | 0.000 (0.000) | -0.001** (0.001) |
| Race | Yes | Yes | Yes | Yes | Yes | Yes | Yes | Yes | Yes |
| Industry | Yes | Yes | Yes | Yes | Yes | Yes | Yes | Yes | Yes |
| Residual standard error (df) | 16.27 (13,879) | 16.86 (5,969) | 15.3 (2,737) | 0.850 (13,878) | 0.860 (5,968) | 0.829 (2,736) | 0.848 (13,878) | 0.872 (5,968) | 0.812 (2,736) |
| Multiple R-squared (Adjusted) | 0.118 (0.116) | 0.109 (0.103) | 0.107 (0.094) | 0.183 (0.181) | 0.237 (0.231) | 0.100 (0.082) | 0.089 (0.087) | 0.110 (0.104) | 0.092 (0.078) |
| F-statistic (df) | 45.33*** (41, 13,879) | 17.81*** (41, 5,969) | 8.023*** (41, 2,737) | 74.24*** (42, 13,878) | 44.06*** (42, 5,968) | 6.91*** (42, 2,736) | 32.47*** (42, 13,878) | 17.59*** (42, 5,968) | 6.568*** (42, 2,736) |

*Figure 7 Predicted productivities by sex across firms of different long-term policy on WFH (Reference group: Female)*

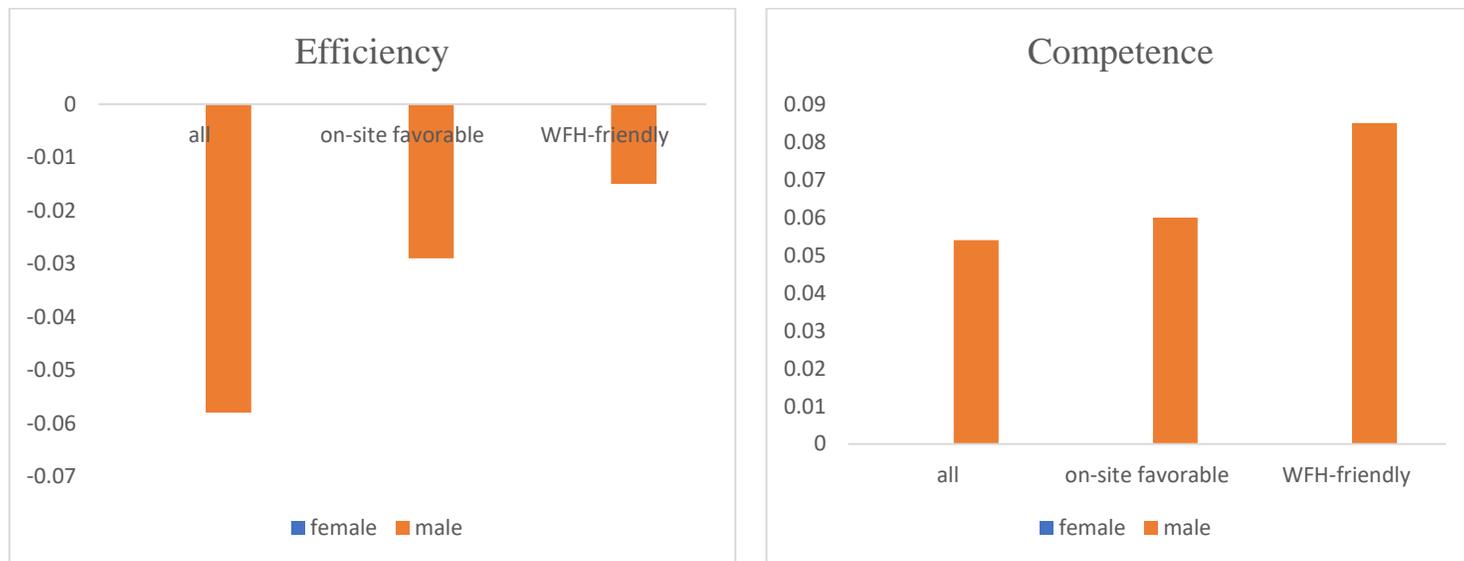

*Figure 8 Predicted productivities by race across firms of different long-term policy on WFH (Reference group: Black or African American)*

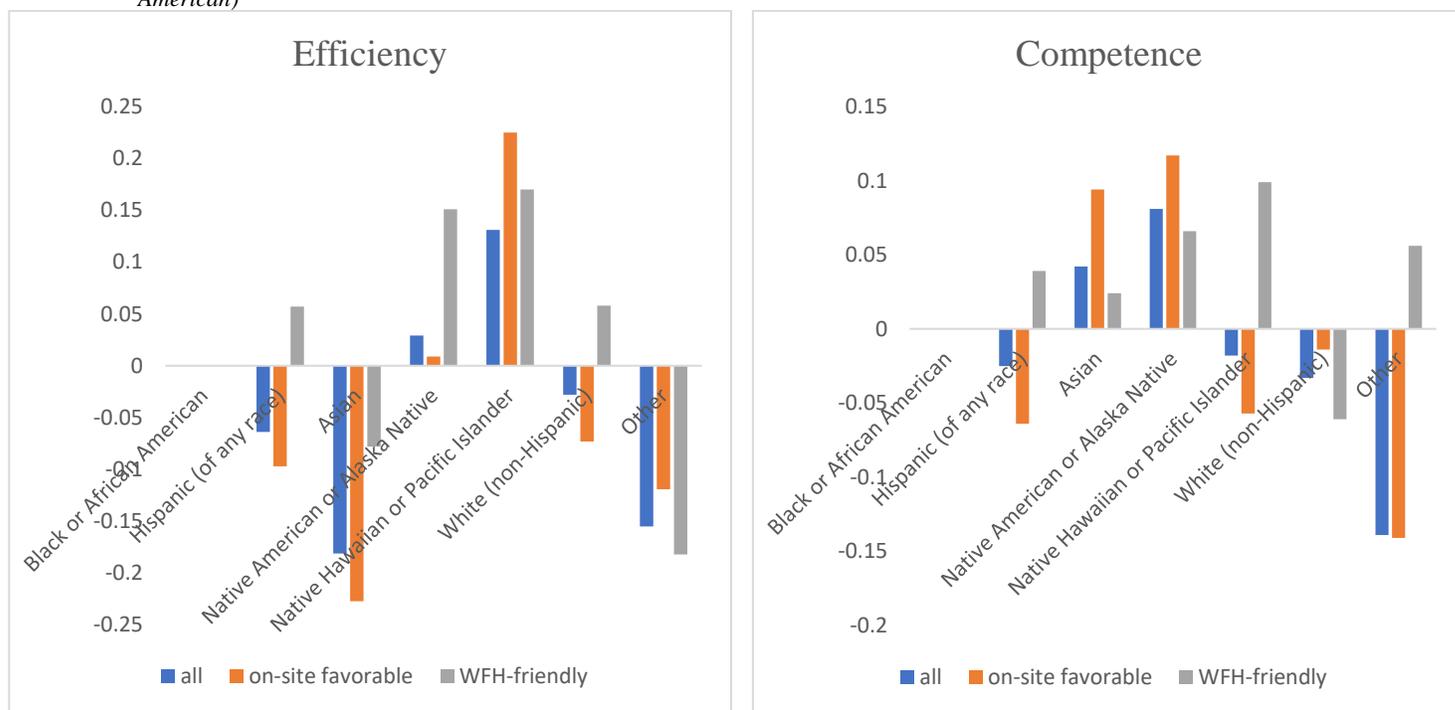

*Figure 9 Predicted productivities by industry across firms of different long-term policy on WFH (Reference group: Agriculture)*

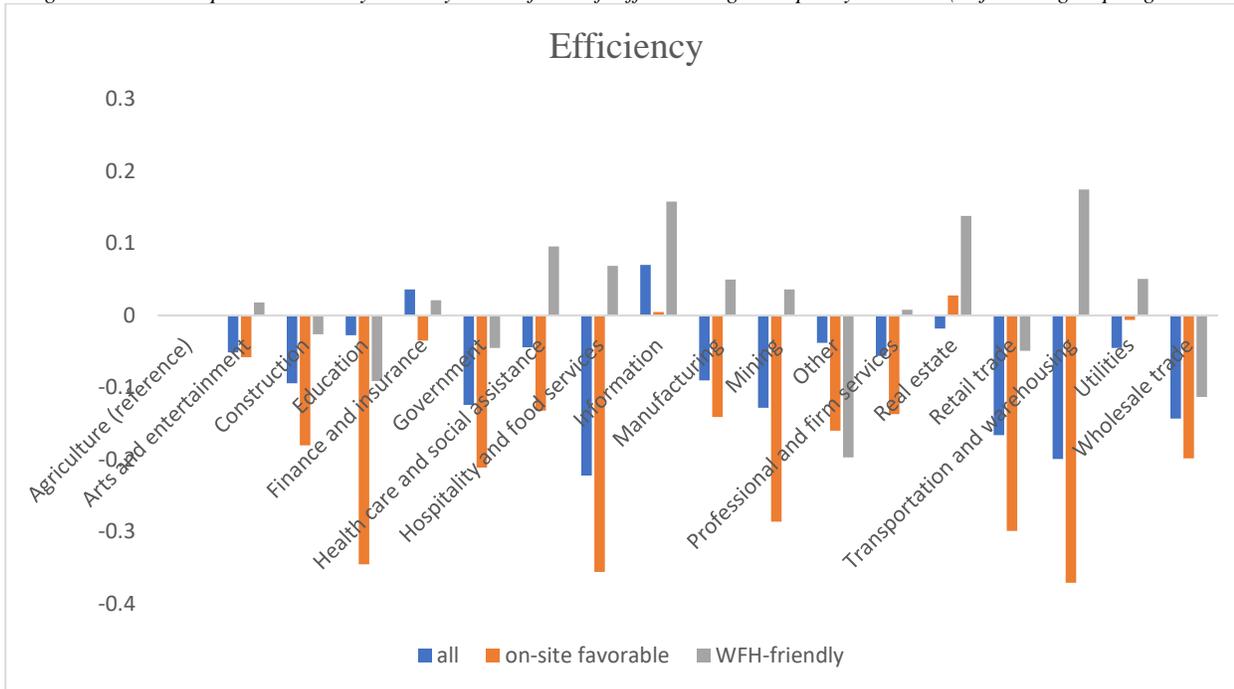

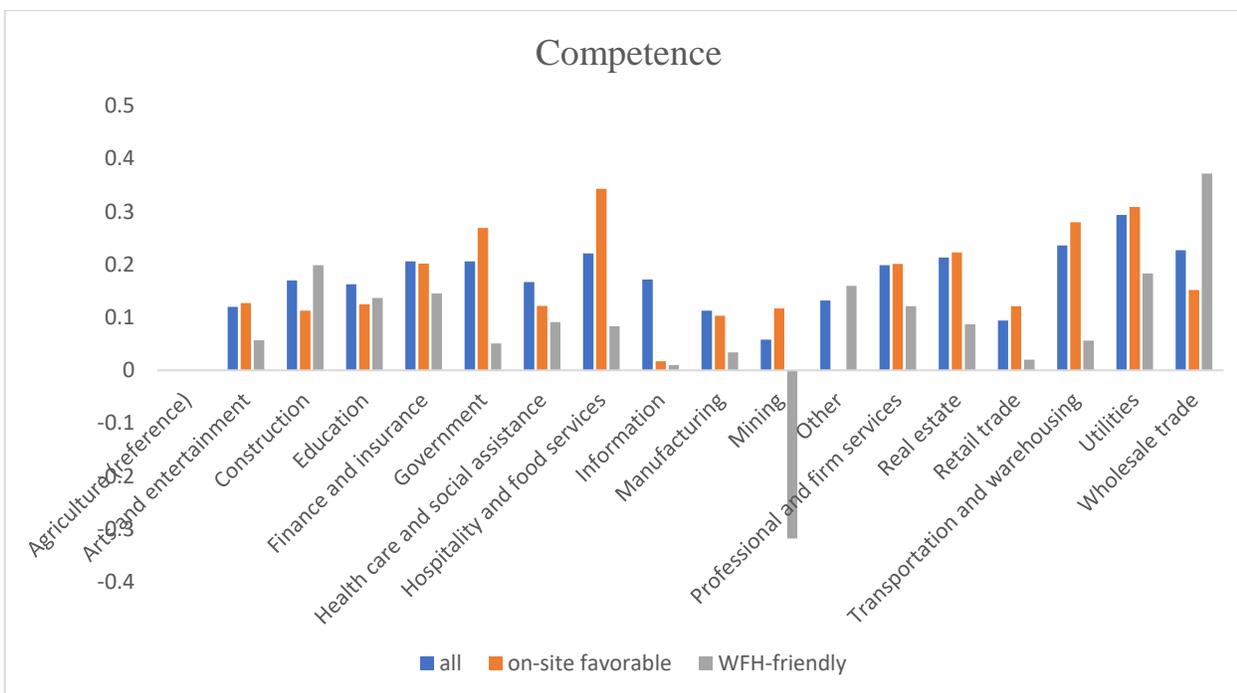

to implement moral hazard while it is possible. These findings suggest that firms differentiate their strategies on WFH policy according to the type of productivities they seek for the business: (1) If efficiency mainly explains firm's business purpose, the permanent WFH policy would block workers from implementing moral hazard; (2) Or if competence matters the most for the business, any types of WFH policy, either temporary or permanent, would end up yielding workers' moral hazard. Thus, a set of incentivized systems specifically targeted and designed for the WFH environment needs to be developed so as to reduce the potential risk of workers' moral hazard.

  The results also suggest a great advantage of WFH for workers with health issues. Even though remote work may favor disabled workers with respect to accessibility, disabled workers had not had opportunities to work from home before pandemic (Holland, 2021). However, COVID 19 have provided them with opportunities to work from home and demonstrated that it is possible (Brown et al., 2021). As seen in Table 5, those who have a health issue or a disability that prevents work or limits the kind or amount of work showed significantly greater efficiency and competence than those who do not. This implies that WFH environment could be even more supportive and productive workplace for workers with health issues partly because they can work with fewer constraints. While studies on health issues in the WFH culture argue its hindering impact on health (Oakman et al., 2020; Ekpanyaskul and Padungtod, 2021; Xiao et al., 2021), the finding in this study suggests beneficial aspects of WFH regarding health issues, particularly on the disabled workers.

  Additionally to the previous literature about conditions to be effective in the WFH environment (Gadeyne, 2018; Galanti, 2020; Galanti, 2021), this study finds that making an effort to be effective in a WFH environment is required for both the firm and worker and working conditions at home, such as IT capital quality, at-home offices, and childcare by a partner or others, must be met. Another challenge for the WFH environment is that the traditional team collaboration system via virtual tools may not help increasing efficiency or competence in the WFH environment. Still, teamwork collaboration is worth pursuing since workers' preference regarding interactions with coworkers or clients turns out to have positive effect on competence as found in Table 5. It calls for firms to pay attention to team collaboration skills specifically motivated and designed for the WFH culture, acquainted with studies on successful virtual communications introduced expressly for this purpose (Morrison-Smith and Ruiz, 2020; Feitosa and Salas, 2021; Feitosa et al., 2021; Reyes et al., 2021; Aquino et al., 2022).

## 5. Conclusion

Triggered by a pandemic, an enforced WFH situation opened a new phase for the working environment. The structural dissimilarity between in-person and remote work has induced major anxiety for firms as they are not able to interact with or manage workers in the same way as before the pandemic, which may lead to worker's undermined productivity. Increasing sales in the employee-monitoring software programs may signal firms' fears.

The belief that workers' productivity is manageable through monitoring indicates that productivity has a simple structure and can be controlled with surveillance. However, the conceptual and empirical evidence for the multidimensional nature of productivity urges its diverse and multifarious aspects. This implies that workers' productivities may not be fully subject to monitoring. To predict workers' productivities underlying their multidimensional productivities, the representative latent factors were respectively labeled as *efficiency* and *competence* in productivity and featured by the loadings on a specific set of productivity components.

The key finding is that the workers' ex-post moral hazard induced by the unexpected shock on the firms' value is found to be significant except for the workers at firms with WFH-friendly policy for long term. This finds how continuity in WFH policy discourages workers to implement moral hazard in terms of efficiency, whereas an incentivized system on wages specifically designed for workers' competence generally needs to be developed. Moreover, the results show the advantages for workers with a health issue or a disability whose efficiency and competence are greatly enhanced. The conditions for workers to have effective productivities in the WFH environment include both firm & worker's effort on WFH culture and supportive working conditions at home. Team collaboration seems to be a main challenge for WFH culture, which calls for effective learnings and practices in virtual communications.

Altogether, this study stresses the importance of comprehending workers' productivities in the various aspects that could not be totally monitored by surveillance and proves an empirical evidence of the ex-post moral hazard among the current WFH workers. Findings in this study reveal the strengths and weaknesses of the current WFH working conditions and may help firms decide which policy they may take on WFH culture depending on their businesses and situations. Due to the unavailability of data, however, a limitation of this study is that each worker is assumed to be representative of each firm. This limitation could be improved with some available data that fits this assumption or other methodology that could overcome it.

**Appendix**

*Table A 1 Variables and definitions*

| Variable codes | Definition |
|---|---|
| employer_arr_qual | What plans does your Firm have for working arrangements of full-time employees after COVID, in 2022 or later?<br>1 Fully on-site<br>2 Hybrid: 1 to 4 days WFH<br>3 Fully remote |
| *Firm's value* | |
| Workteam _npeople | How many people belong to your main work team? (Top-coded at 50) |
| *Worker's productivity components* | |
| wfh_eff_COVID_quant | How efficient are you WFH during COVID, relative to on business premises before COVID? (%) |
| wfh_expect_quant | Relative to expectations before COVID, how productive are you WFH during COVID? (%) |
| prom_eff_1day_quant | How much of an increase in your chance of a promotion would working from home one more day per week than your co-Workers cause? |
| prom_eff_5day_quant | How much of an increase in your chance of a promotion would working from home 5+ days a week while your co-Workers work on the business premises 5+ days a week cause? |
| wfh_extraeff_comm_ quant | How much of your extra efficiency when working from home is due to the time you save by not commuting? |
| extratime_1stjob | Percent of commute time savings spent working on primary or current job |
| *WFH-related variables* | |
| wfh_hoursinvest | Hours invested in learning how to WFH effectively |
| wfh_invest_burs | Percent of money invested in equipment or infrastructure enabling WFH that was paid for or reimbursed by Firm |
| numwfh_days_ postCOVID _boss_s_u | Firm planned share of paid working days WFH after COVID (%) |
| numwfh_days_ postCOVID _boss_pre | Firm planned share of paid working days WFH after COVID, before the most recent announcement made in the past 6 months (%) |
| videocalls_ current_percent | Currently, what percentage of your normal working day do you spend in video calls? |
| videocalls_ preCOVID _percent | Before the pandemic, what percentage of your normal working day did you spend in video calls? |
| wfh_ownroom _notbed 100 | 100 x 1(Has their own room (not bedroom) to work in while WFH during COVID) |
| hours_cc_partner | Currently, how many hours of childcare each week are provided by your partner? |
| hours_cc_other | Currently, how many hours of childcare each week are provided by your partner or by others, e.g. grandparents, babysitters? |
| internet_quality _quant | Internet quality - Fraction of time that internet works |
| disability_qual | Do you have a health problem or a disability which prevents work or which limits the kind or amount of work you do? |
| | 1 Yes |
| | 2 No |
| | 3 Prefer not to answer |
| live_children | Do you currently live with children under 18? -- categorical by youngest's age |
| workteam_tasks _percent | To perform your job, what percentage of your tasks require collaboration as part of a team? |
| coworker_interactions | How much do you enjoy your personal interactions with coworkers at your Firm's worksite? |
| client_interactions | How much do you enjoy your personal interactions with customers, clients, or patients at your Firm's worksite? |
| *Demographic covariates* | |
| gender | 1 Female |
| | 2 Male |
| | 3 Other/prefer not to say |
| race_ethnicity | 1 Black or African American |
| | 2 Hispanic (of any race) |
| | 3 Asian |
| | 4 Native American or Alaska Native 5 Native Hawaiian or Pacific Islander |
| | 6 White (non-Hispanic) |
| | 7 Other |
| educ_years | Years of education |
| age_quant | Age in years |
| Age2 | Age squared |
| *Control variables* | |
| work_industry | 1 Agriculture |
| | 2 Arts and Entertainment |
| | 3 Finance and Insurance |
| | 4 Construction |
| | 5 Education |
| | 6 Health Care and Social Assistance |
| | 7 Hospitality and Food Services |
| | 8 Information |
| | 9 Manufacturing |
| | 10 Mining |
| | 11 Professional and Business Services |
| | 12 Real Estate |
| | 13 Retail Trade |
| | 14 Transportation and Warehousing |
| | 15 Utilities |
| | 16 Wholesale Trade |
| | 17 Government |
| | 18 Other |
| commutetime_quant | Commute time (mins) |

*Validity of Self-Reported Productivity Measures*

Because of the unavailability of direct or objective measures in various situations, social science depends heavily on self-reported measures. However, according to the *American Psychological Association Dictionary of Psychology*, there is widespread concern about self-report bias. Respondents' thoughts, feelings, or behaviors may not be as accurate as the direct or objective measures. Respondents may not fully understand the questions, they may answer as is socially desired to give a good impression, or they may be confused about what they feel.

Nonetheless, empirical evidence supports the validity of self-reports. For example, Allen and Bunn (2003) conducted joint analyses of adverse event measures and self-reported productivity measures on employees at International Truck and Engine Corporation. They found self-reported measures to perform well in concurrent and predictive validities. In addition, Brener et al. (2003) reviewed over 100 studies investigating the validity of self-reported health-risk behaviors such as alcohol, tobacco, and other drug use as well as unintentional injuries and violence, dietary, physical, and sexual behaviors among adolescents. Even if those health-risk behaviors were affected by cognitive and situational factors, the validity of self-report measures was not threatened as long as the research was designed to minimize threats to inherent validity as much as possible. The Center for Health and Safety Culture (2011) argued for the validity of self-report survey data, given that the respondents understand the questions and the conditions of a strong sense of anonymity and a lack of fear of reprisal. Accordingly, the validity and accuracy of self-reported productivity measures are assumed to be satisfied with well-designed surveys.

*A summary for factor analysis*

This summary-for-factor analysis is primarily from the lecture notes (Park, 2021). Consider a multivariate normal distribution of $(X, Y)$ with

$$(X, Y) \sim N(M(\phi, \lambda), I_n \otimes \Sigma), \quad \Sigma = \begin{pmatrix} \Sigma_{xx} & \Sigma_{xy} \\ \Sigma_{yx} & \Sigma_{yy} \end{pmatrix},$$

where $Y$ is $n \times q$ observed; $X$ is $n \times p$ latent or unobserved; $M$ is $n \times k$ known; and $\phi, \lambda$ and $\Sigma$ are $k \times p, k \times q$, and $q \times q$, respectively, and unknown. Then, $Cov(Y|X) = I_n \otimes \Sigma_{yy \cdot x} = I_n \otimes (\Sigma_{yy} - \Sigma_{yx}\Sigma_{xx}^{-1}\Sigma_{xy})$ if $\Sigma_{xx}$ is nonsingular. The essential assumption for factor analysis is that the observed variables $Y$ are orthogonal conditional on the latent factors $X$. Accordingly, I write $\Sigma_{yy \cdot x}$ as a diagonal matrix

$$\Sigma_{yy \cdot x} = \Omega = \begin{pmatrix} \omega_{11} & \cdots & 0 \\ \vdots & \ddots & \vdots \\ 0 & \cdots & w_{qq} \end{pmatrix},$$

where the respective unique variance (uniqueness) of variable $j$, $\omega_{jj}$, is nonnegative for all $j$. Note that the unique variances are the leftover variability of each variable that is not explained by the factors. With this assumption, the marginal distribution of $Y$ is given by

$$Y \sim N\left(M\lambda, I_n \otimes \left(\Sigma_{yx}\Sigma_{xx}^{-1}\Sigma_{xy} + \Omega\right)\right).$$

A conventional choice is $\phi = 0$ and $\Sigma_{xx} = I_p$ because it is well known that the distribution of $Y$ does not depend on the mean of $X$ and does not change by an affine transformation of $XA$, for invertible $A$. Thus

$$Y \sim N\left(M\lambda, I_n \otimes (\beta'\beta + \Omega)\right),$$

letting $\beta = \Sigma_{xx}^{-1}\Sigma_{xy} = \Sigma_{xy}$. Remark that the factor analysis model assumes a particular decomposition of the covariance matrix, which is different from the ordinary least squares model. The parameters of interest are $\beta$ and $\Omega$ to be estimated via the maximum likelihood estimation, based on $U = Y'Q_M Y$, which follows the Wishart distribution

$$U \sim W_q(\upsilon = n - k, \ \beta'\beta + \Omega),$$

and $Q_M = I_n - P_M$, with $P_M$ being a projection onto the column space of $M$. The identifiability for $\beta$ is subject to the inequality of

$$(q - p)^2 - p - q \geq 0. \qquad \text{(A1)}$$

Under the regular restrictions on the parameters, $\Sigma_{yy} = \beta'\beta + \Omega$ is unique for most $\Sigma_{yy}$ with little exceptions such as $\Sigma_{yy} = I_q$.

The optimal number of factors can be determined by a criterion for model selection such as the Akaike information criterion (AIC; Akaike, 1974) or the Bayesian information criterion (BIC; Schwarz, 1978). Both are popular likelihood-based information criteria, but each is preferred for a different purpose:AIC works better for prediction, and BIC is better for the correct model (Berk, 2008; Konishi and Kitagawa, 2007; Shmueli, 2010; Sober, 2002). Let $M_p$ denote the $p$-factor model. Based on the marginal likelihood, the observed deviance is

$$D\big(M_p\big(\hat{\beta}, \widehat{\Omega}\big); U\big) = v \ln|\hat{\beta}'\hat{\beta} + \widehat{\Omega}| + C,$$

where $C$ is a constant independent of $p$. Then, AIC and BIC are respectively given by

$$\begin{cases} AIC\big(M_p\big) = v \ln|\hat{\beta}'\hat{\beta} + \widehat{\Omega}| + 2d_p & (A2) \\ BIC\big(M_p\big) = v \ln|\hat{\beta}'\hat{\beta} + \widehat{\Omega}| + \ln(v) \, d_p, & (A3) \end{cases}$$

where $d_p = q(p+1) - p(p-1)/2$. The model with the smallest criterion values is selected as the best model, whichever criterion is used.

Moreover, we may further choose a $p \times p$ orthogonal matrix $\Psi$, which gives the best interpretation for $\hat{\beta}^* = \Psi\hat{\beta}$, because $\hat{\beta}^{*'}\hat{\beta}^* = \hat{\beta}'\Psi'\Psi\hat{\beta} = \hat{\beta}'\hat{\beta}$. It is well known that choosing $\Psi$ that maximizes varimax gives a desirable result, where varimax is defined by $\sum_{i=1}^{p} \sigma_i$ with $\sigma_i$ as the variance of the squares of the elements in row $i$ of $\hat{\beta}^*$. Furthermore, from the conditional expectation of $X$ given $Y$ such that

$$E(X|Y = y) = (y - M\lambda)(\beta'\beta + \Omega)^{-1}\beta',$$

the elements of factors can be predicted by

$$\hat{X} = \big(Y - M\hat{\lambda}\big)\big(\hat{\beta}'\hat{\beta} + \widehat{\Omega}\big)^{-1}\hat{\beta}'$$

and used for analysis.

*Figure A 1 Demographic covariates for missing and non-missing variables*

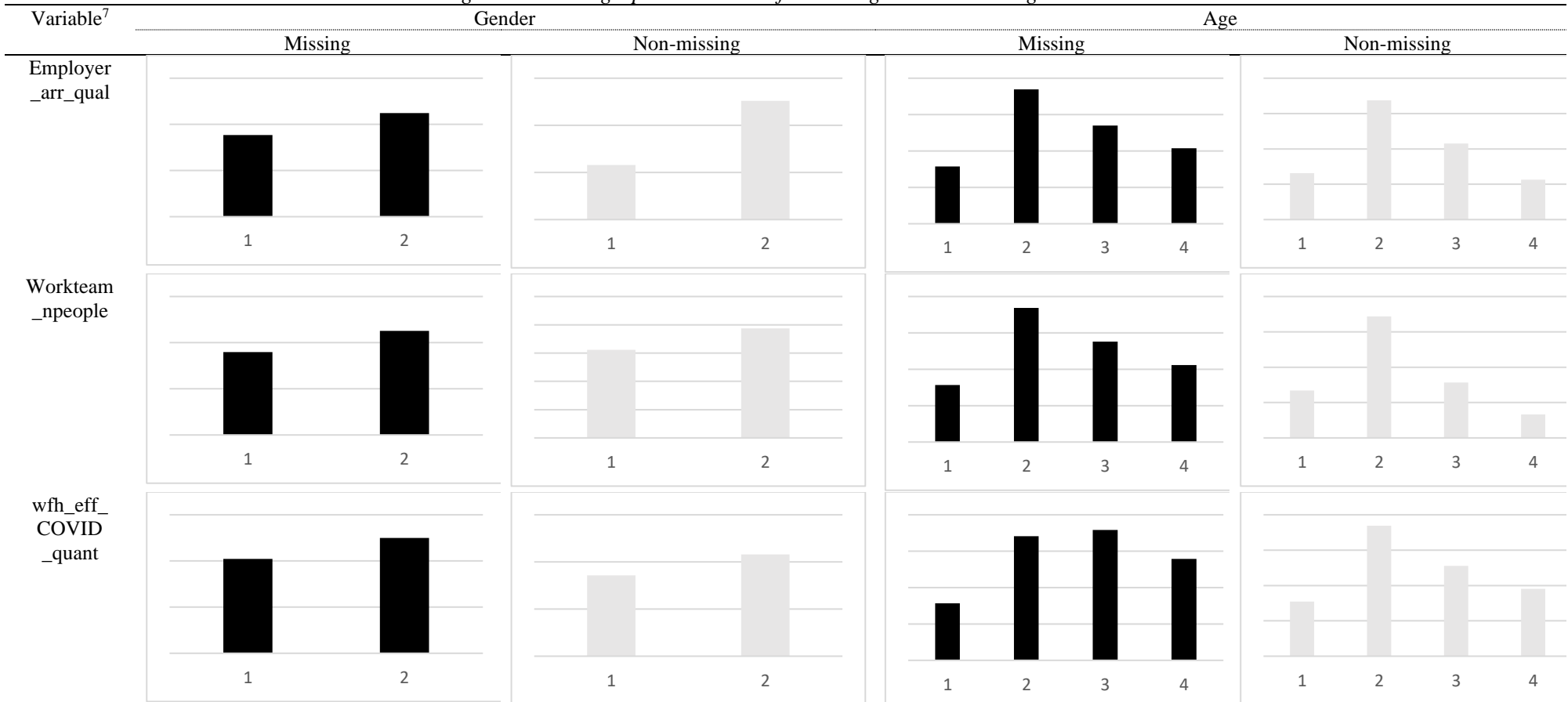

| Variable[7] | Gender | | | | Age | | | |
|---|---|---|---|---|---|---|---|---|
| | Missing | | Non-missing | | Missing | | Non-missing | |

---

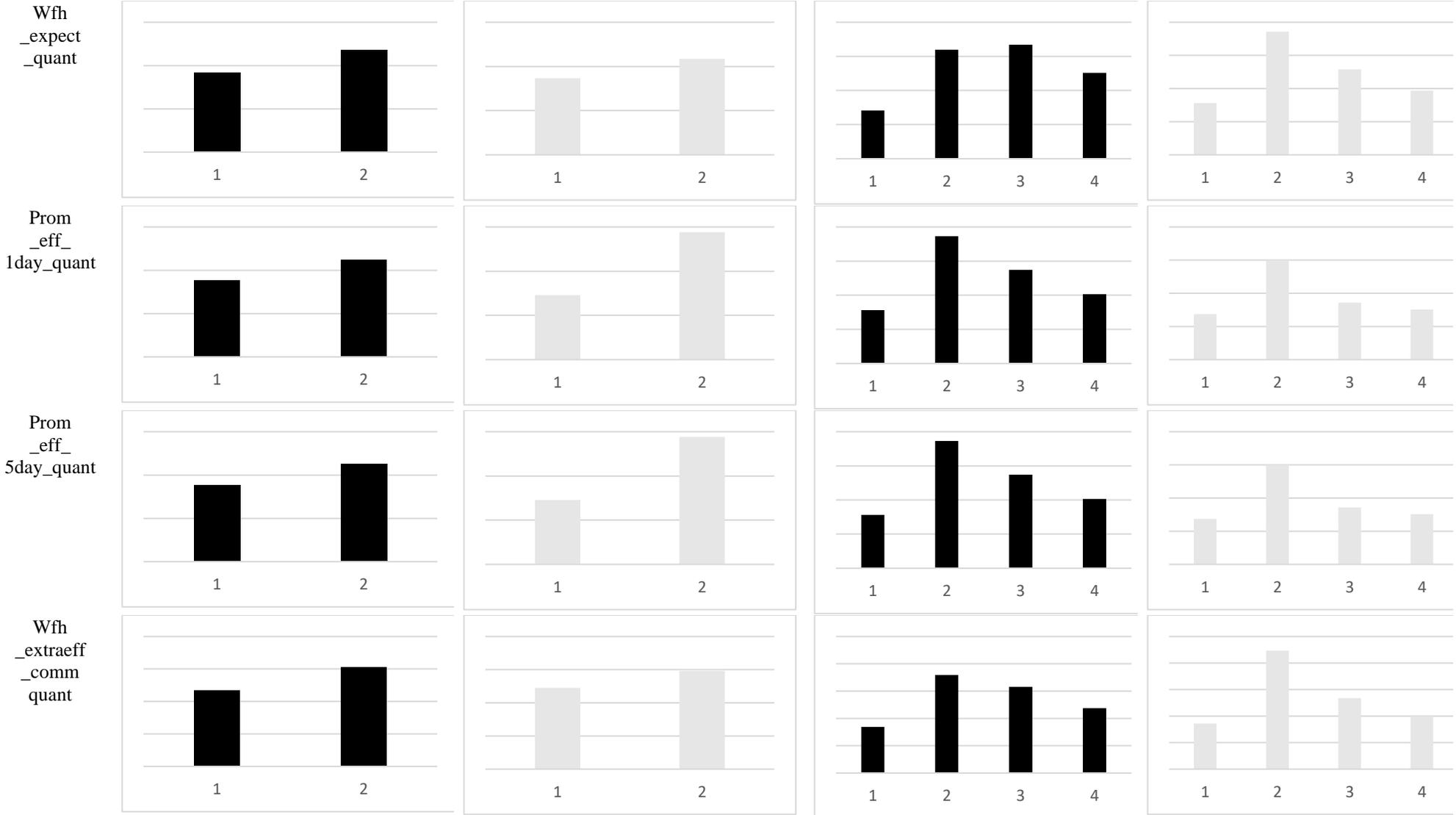

Extratim_
1stjob
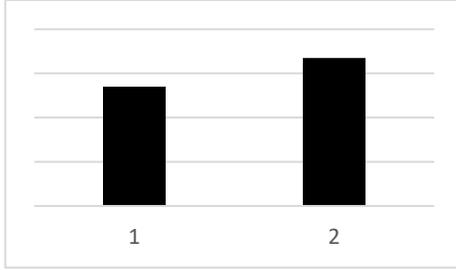 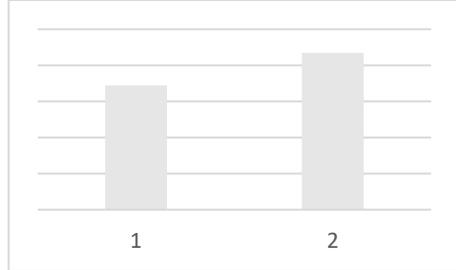 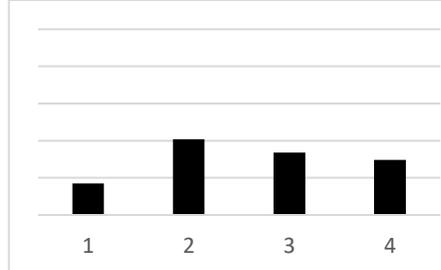 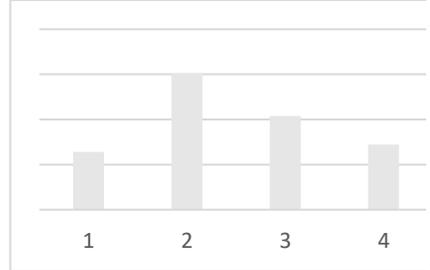

Video
calls_current
_percent
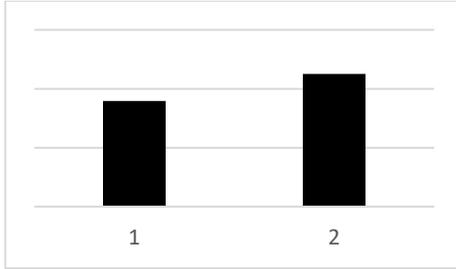 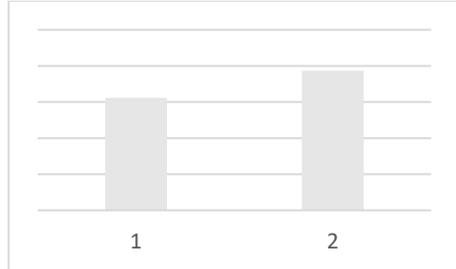 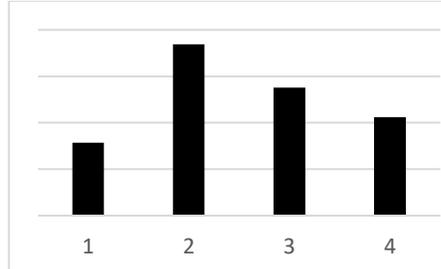 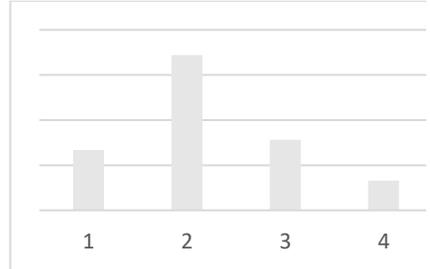

Video
calls_
pre
COVID
_percent
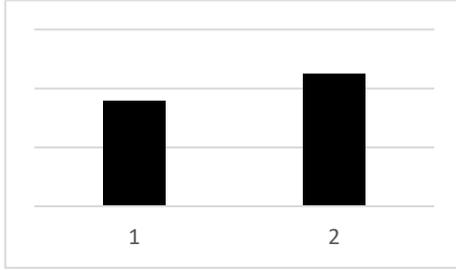 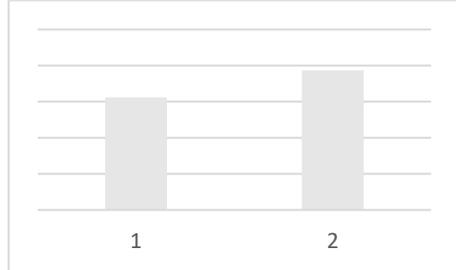 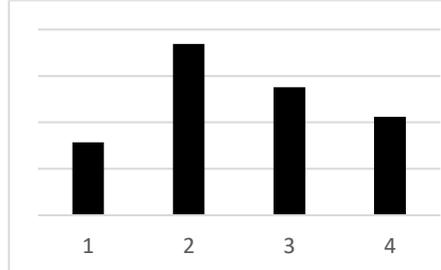 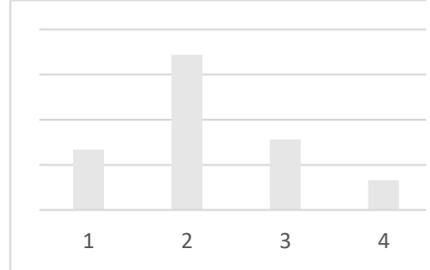

wfh_
invest
_burs
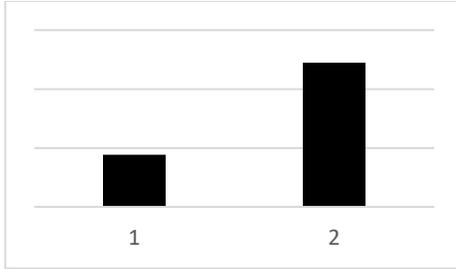 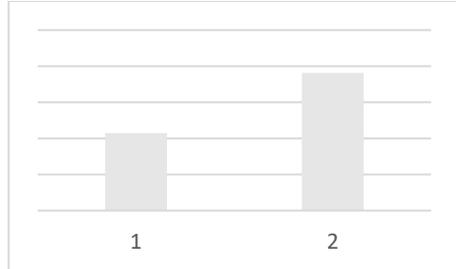 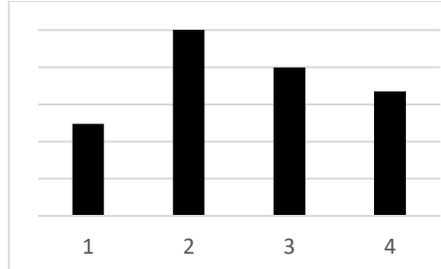 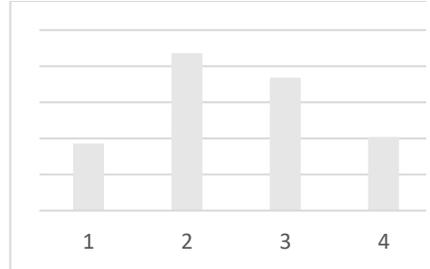

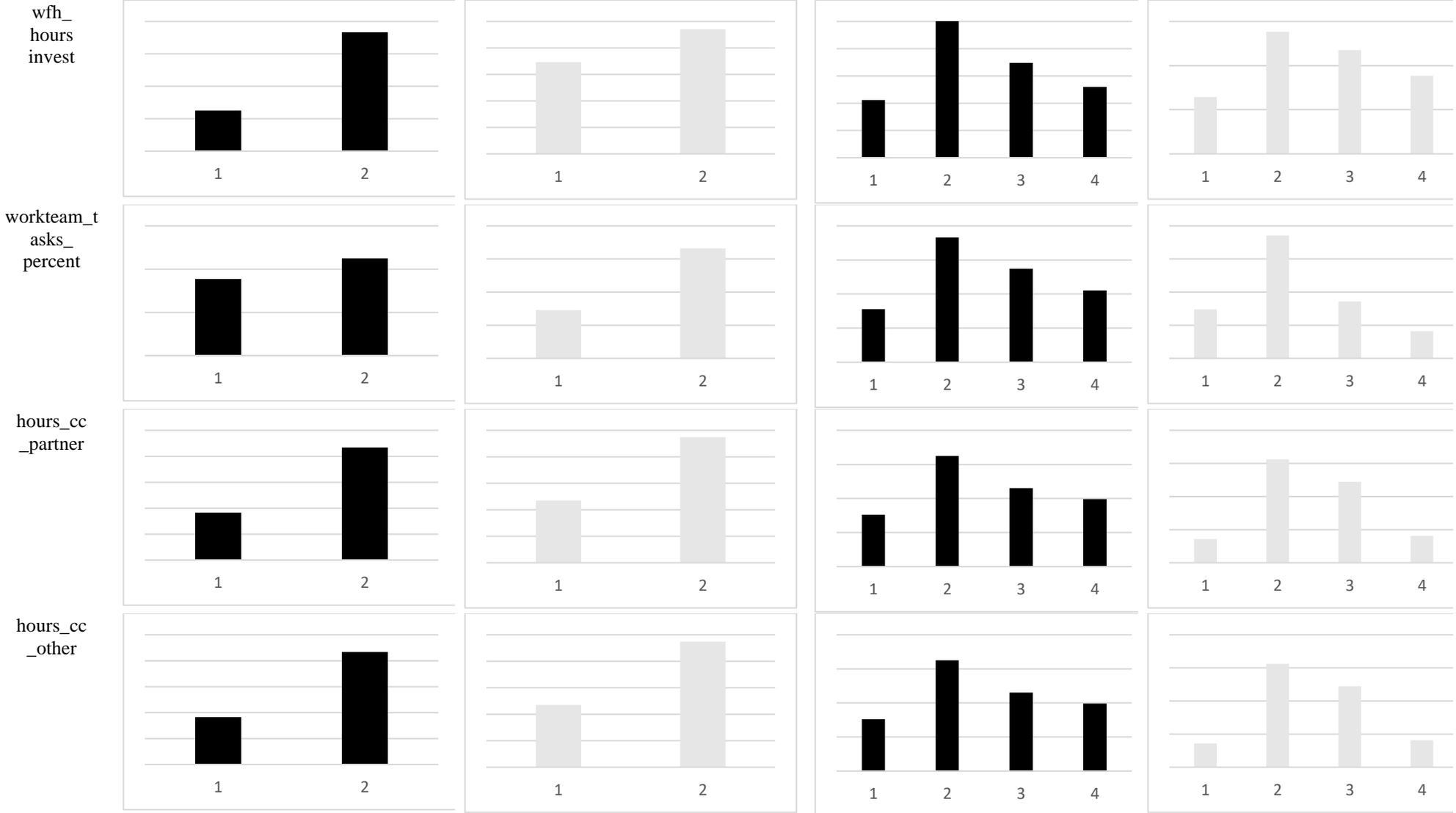

| | | | |
|---|---|---|---|
| numwfh days_ post COVID _boss_pre | 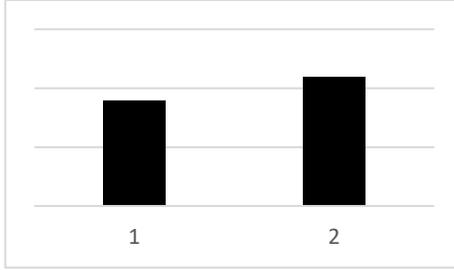 | 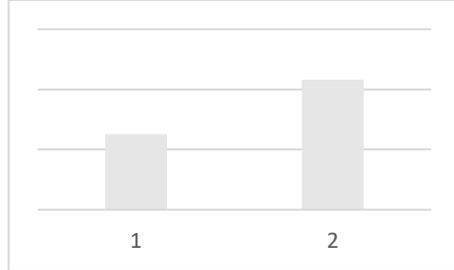 | 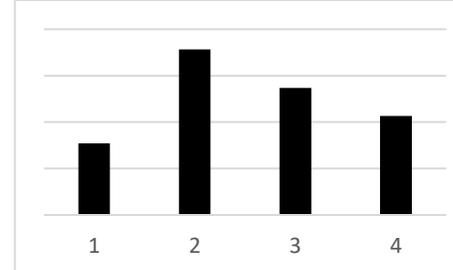 | 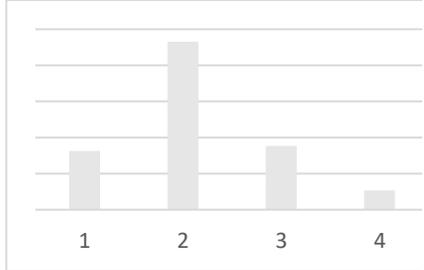 |
| numwfh_days _post COVID _boss_s_u | 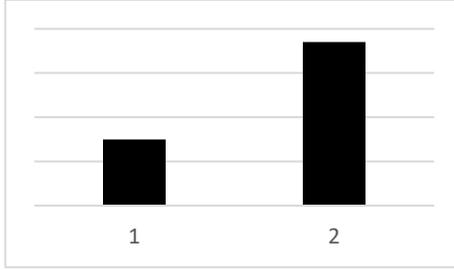 | 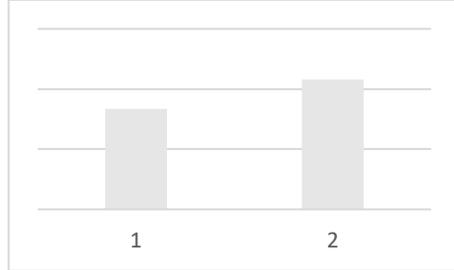 | 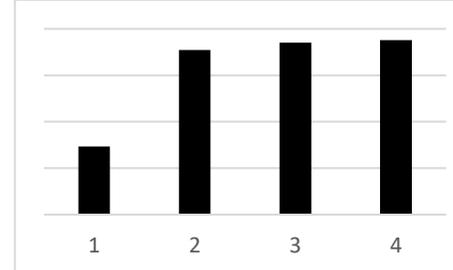 | 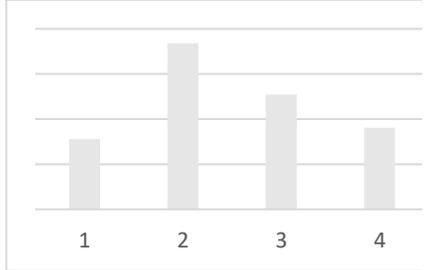 |
| coworker_ inter-actions | 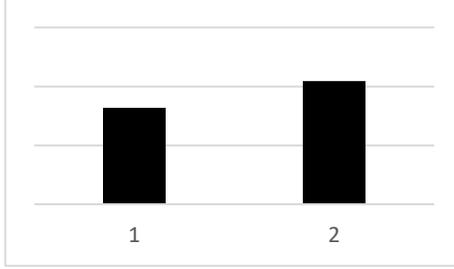 | 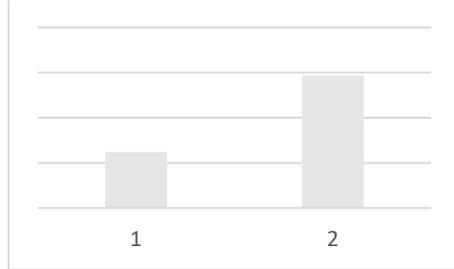 | 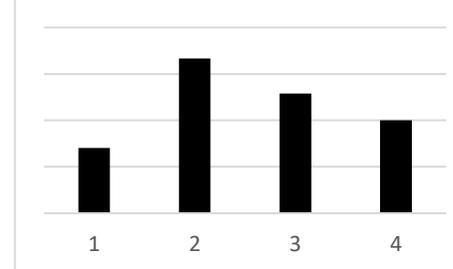 | 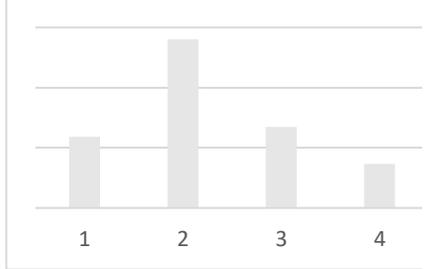 |
| client_ inter-actions | 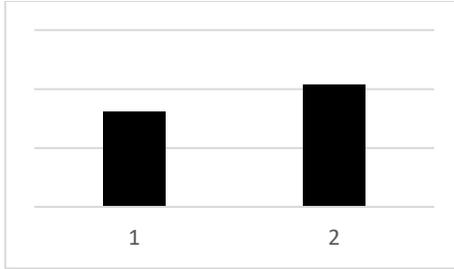 | 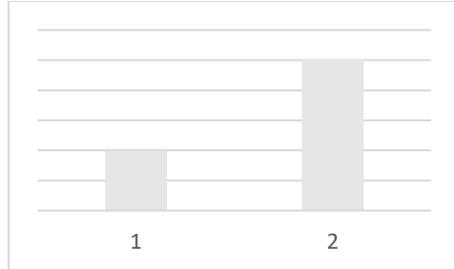 | 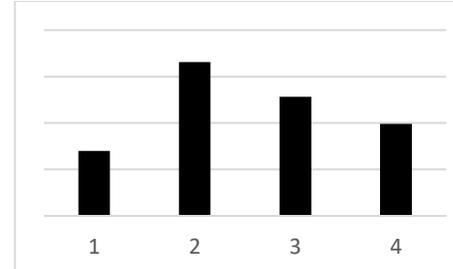 | 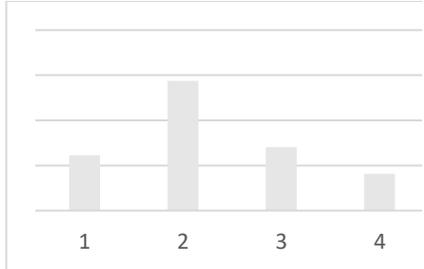 |

| | | | |
|---|---|---|---|
| disability_qual | 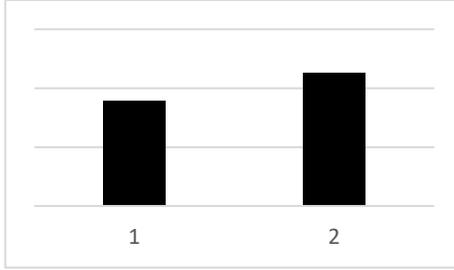 | 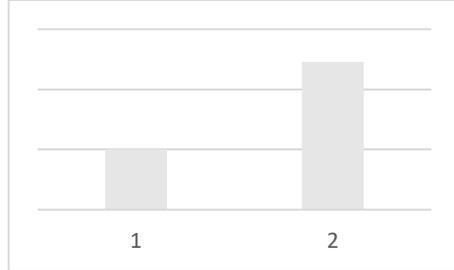 | 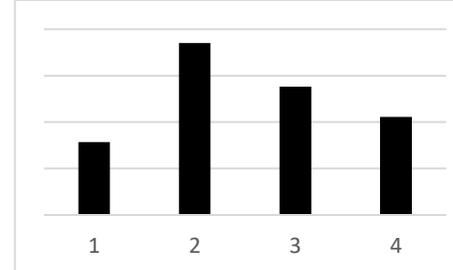 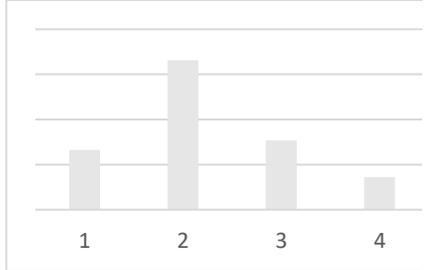 |
| wfh_ownroom_notbed | 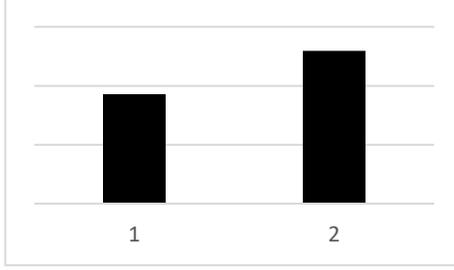 | 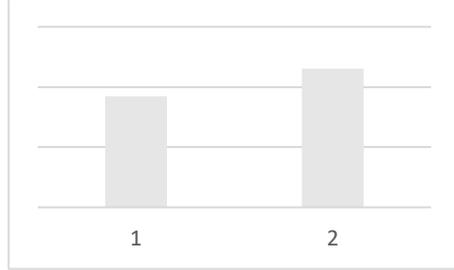 | 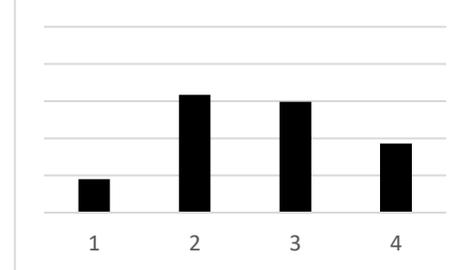 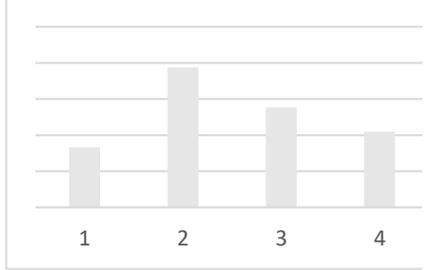 |
| live_children | 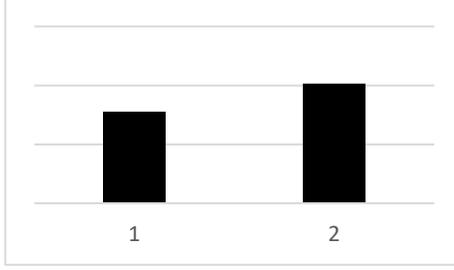 | 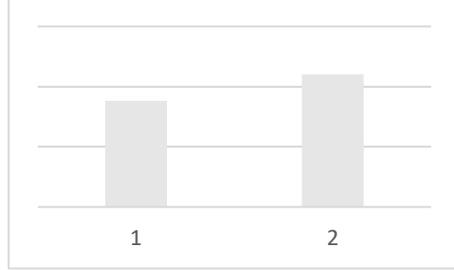 | 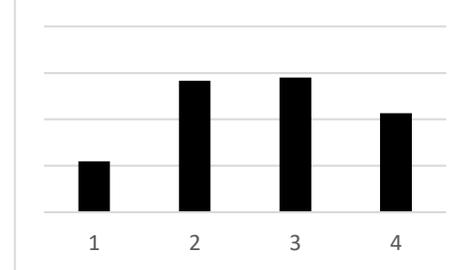 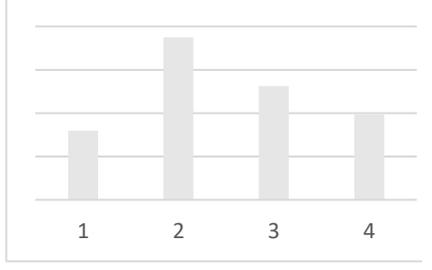 |
| Internet_quality_quant | 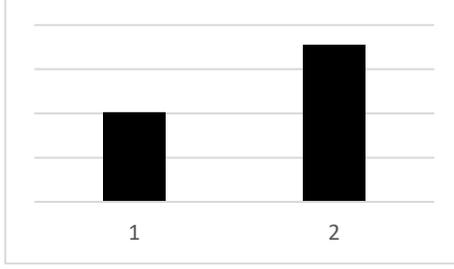 | 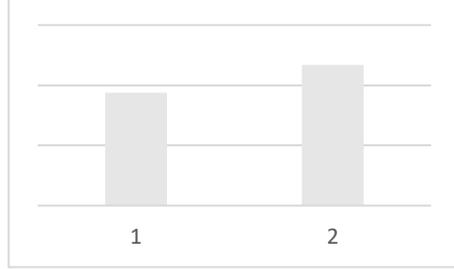 | 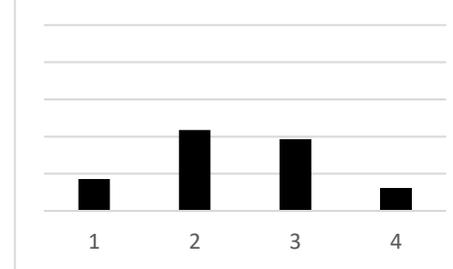 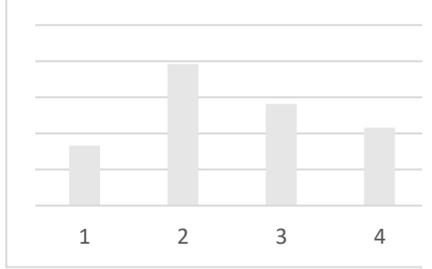 |
| Commute time _quant | 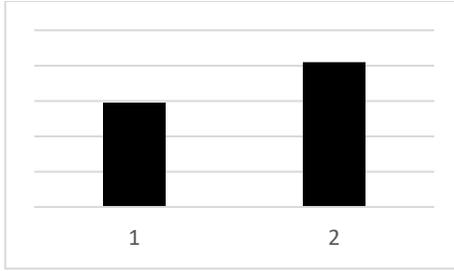 | 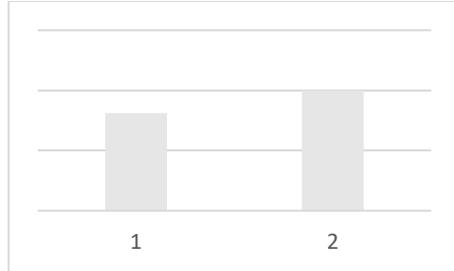 | 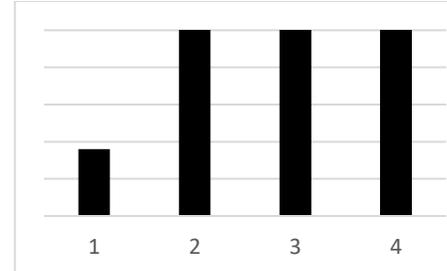 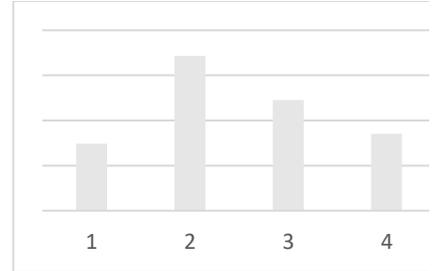 |

| | Years of Education | | Race | |
|---|---|---|---|---|
| Variable | Missing | Non-missing | Missing | Non-missing |
| Employer _arr_qual | 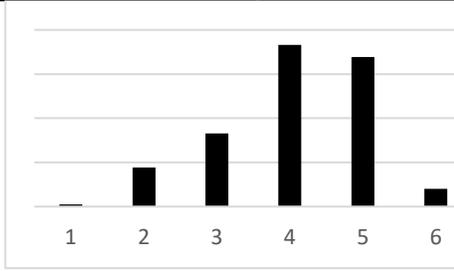 | 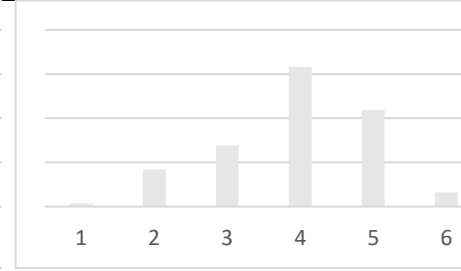 | 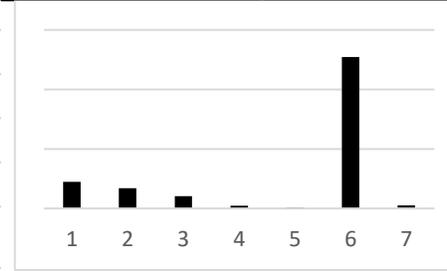 | 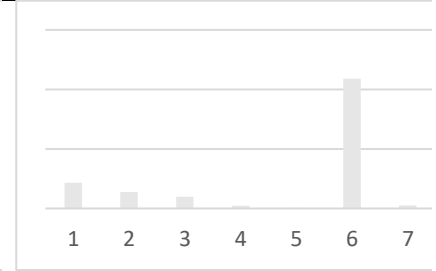 |
| Workteam _npeople | 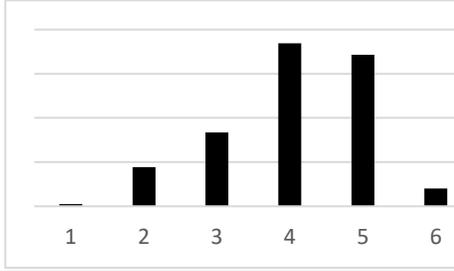 | 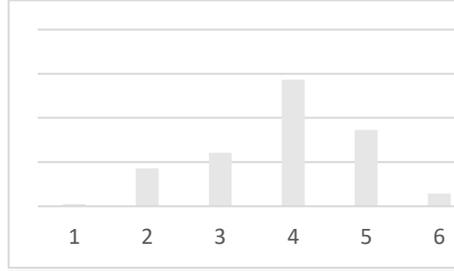 | 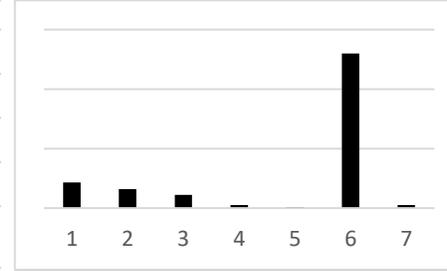 | 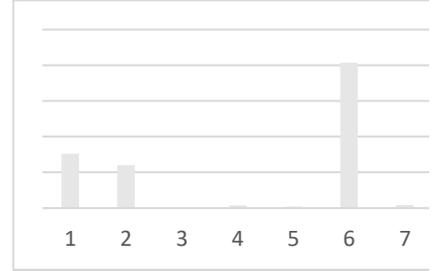 |
| wfh_eff_ COVID _quant | 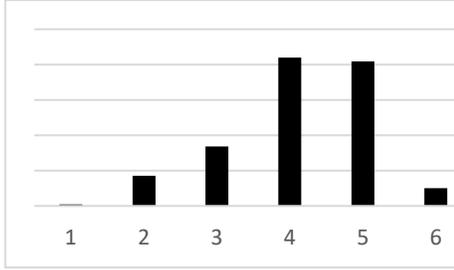 | 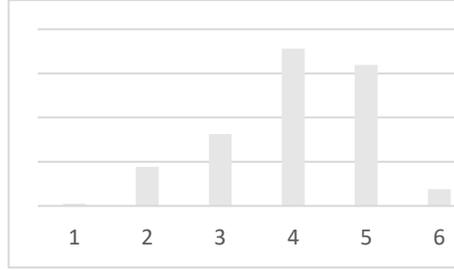 | 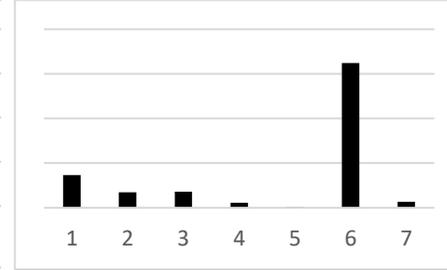 | 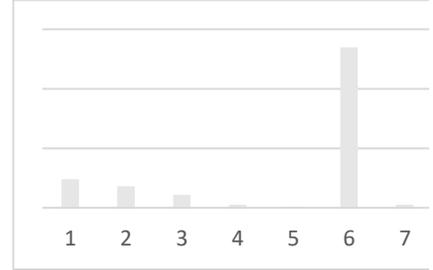 |

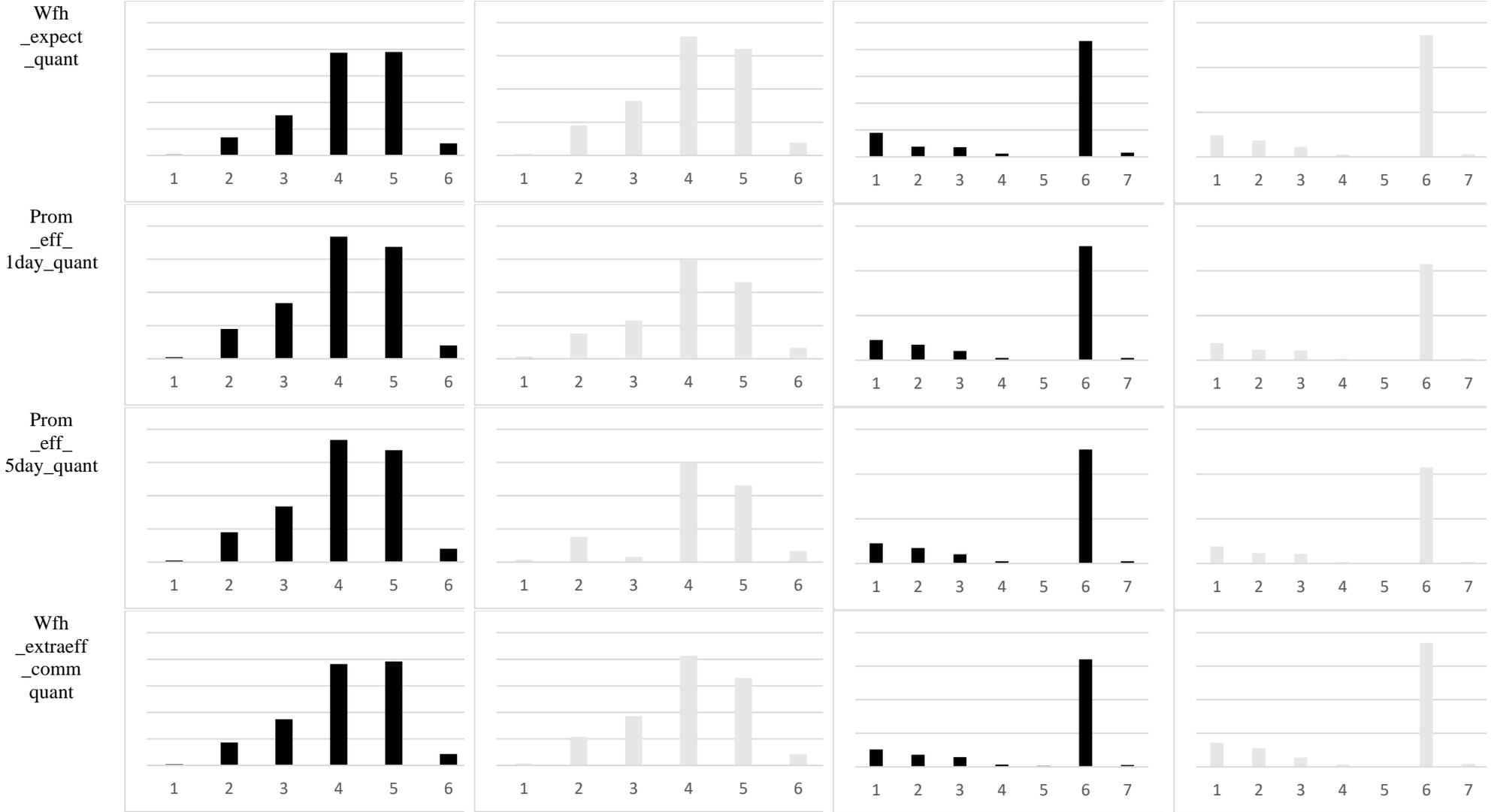

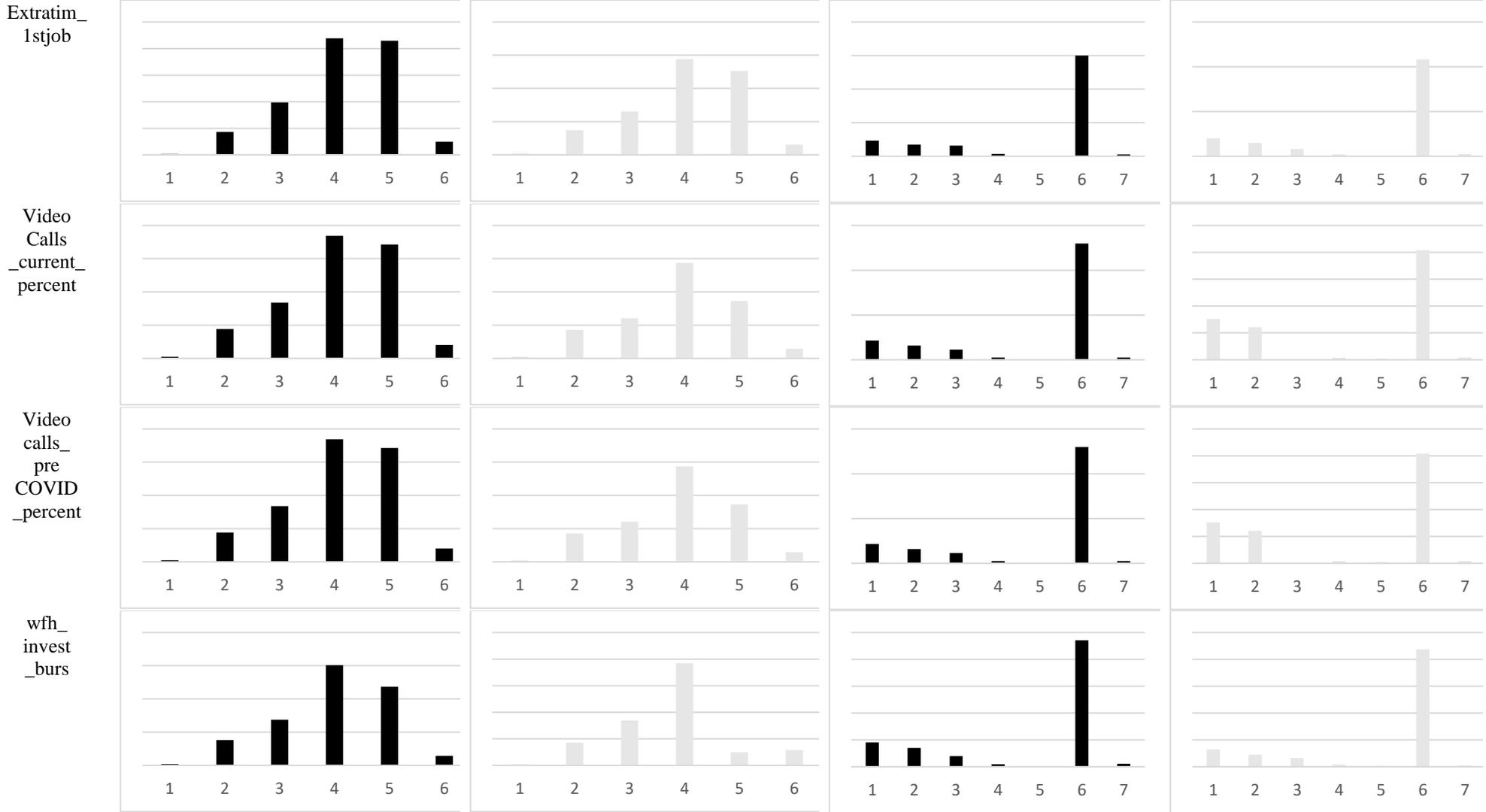

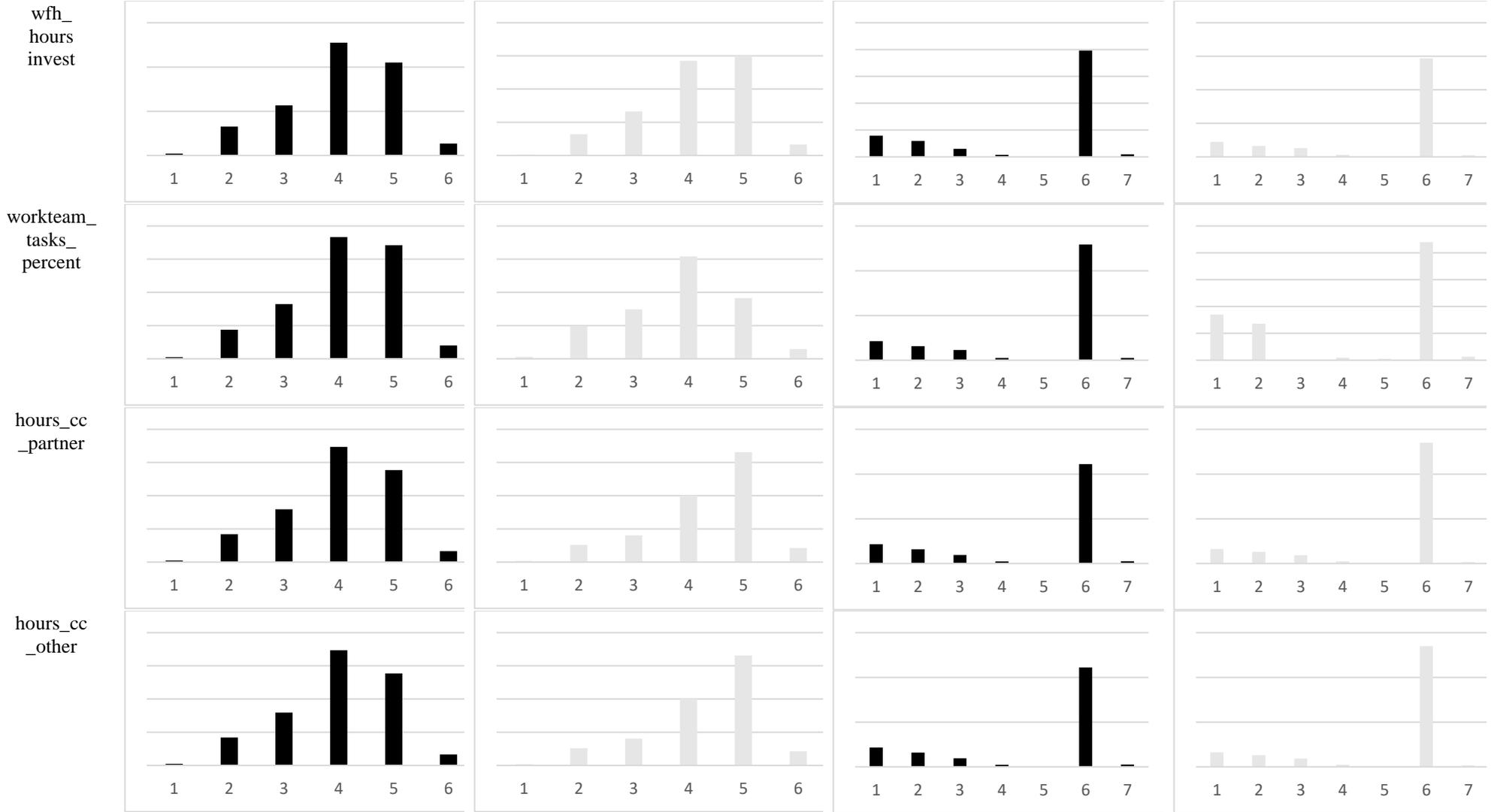

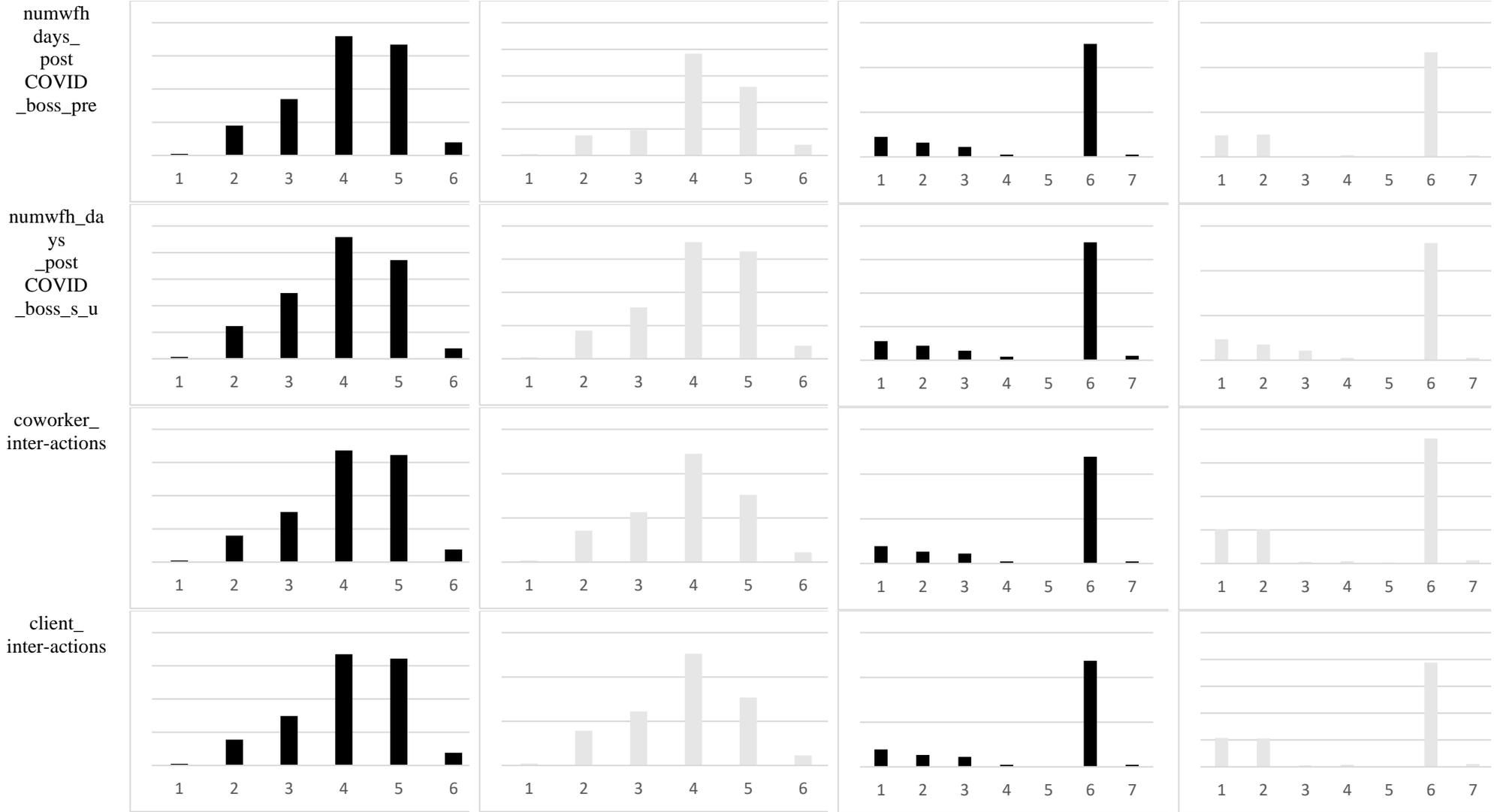

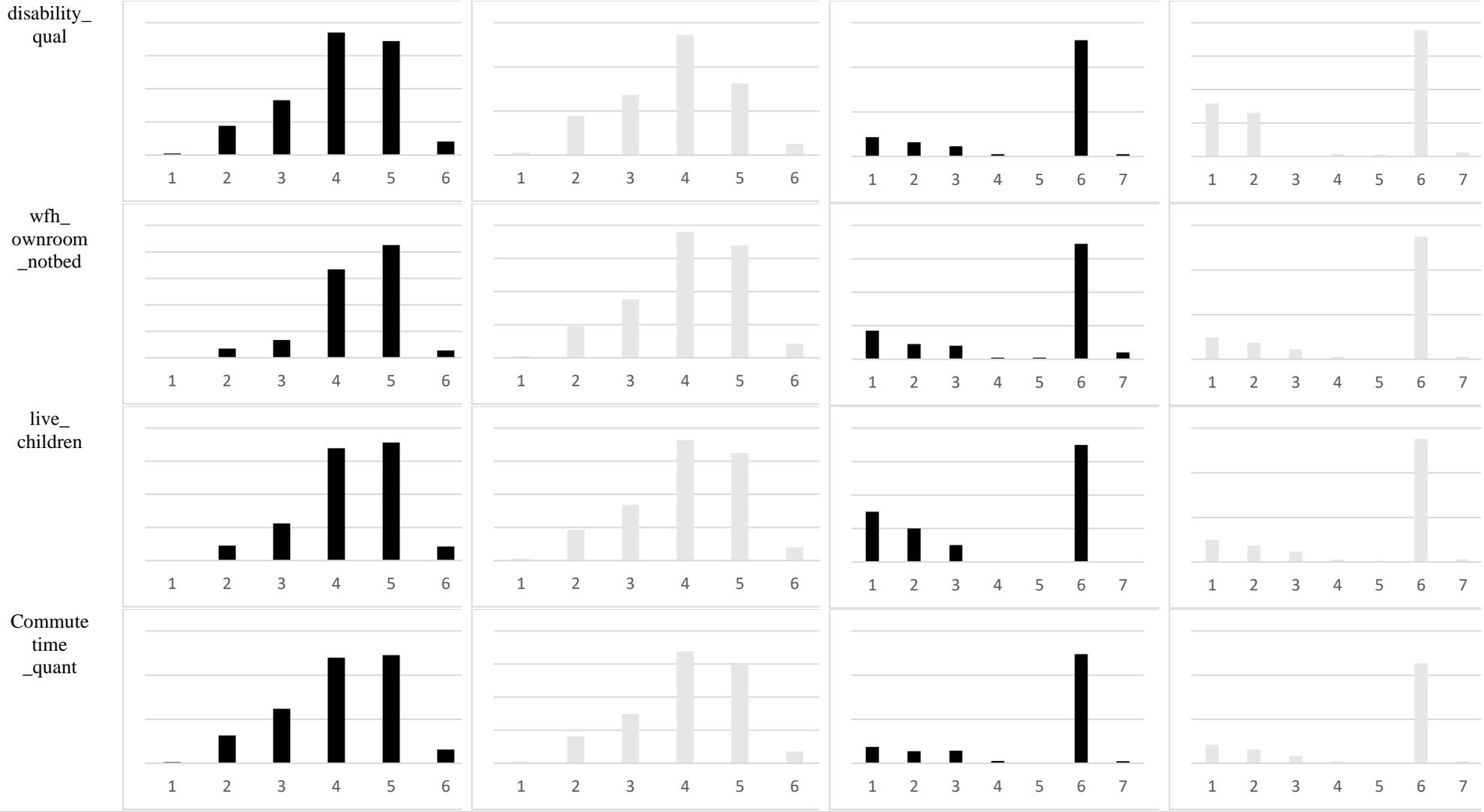

*Table A2 Summary of data for analysis*

| | | All | | | | OnSite-favorable | | | | WFH-friendly | | | |
|---|---|---|---|---|---|---|---|---|---|---|---|---|---|
| 'employer_arr_qual' value | | 1, 2, 3 | | | | 1 (Fully on-site) | | | | 3 (Fully remote) | | | |
| Number of observations | | 13,921 | | | | 6,011 | | | | 2,779 | | | |
| *Continuous variables* | | | | | | | | | | | | | |
| Variables codes | Definition | Mean | St.dev | Min | Max | Mean | St.dev | Min | Max | Mean | St.dev | Min | Max |
| workteam _npeople | How many people belong to your main work team? (Top-coded at 50) | 50.49 | 31.06 | 0.00 | 100.0 | 21.52 | 17.80 | 1.00 | 50.00 | 16.86 | 16.07 | 1.00 | 50.00 |
| wfh_eff_ COVID _quant | How efficient are you WFH during COVID, relative to on business premises before COVID (%) | 10.44 | 17.47 | -50.00 | 40.00 | 11.05 | 18.30 | -50.00 | 40.00 | 9.18 | 16.08 | -50.00 | 40.00 |
| wfh_expect _quant | Relative to expectations before COVID, how productive are you WFH during COVID? (%) | 10.82 | 12.46 | -25.00 | 25.00 | 11.41 | 12.89 | -25.00 | 25.00 | 9.59 | 11.89 | -25.00 | 25.00 |
| prom_eff_ 1day_quant | How much of an increase in your chance of a promotion would working from home one more day per week than your co-Workers cause? | -4.51 | 32.59 | -100.0 | 100.0 | -6.07 | 34.00 | -100.0 | 100.00 | -1.80 | 30.81 | -100.0 | 100.0 |
| prom_eff_ 5day_quant | How much of an increase in your chance of a promotion would working from home 5+ days a week while your co-Workers work on the business premises 5+ days a week cause? | -10.65 | 34.35 | -100.0 | 100.0 | -13.44 | 35.83 | -100.0 | 100.00 | -6.62 | 32.22 | -100.0 | 100.0 |
| wfh_extraeff _comm _quant | How much of your extra efficiency when working from home is due to the time you save by not commuting? | 39.18 | 42.77 | 0.00 | 100.0 | 46.28 | 43.52 | 0.00 | 100.0 | 28.35 | 39.36 | 0.00 | 100.0 |
| extratime_ 1stjob | Percent of commute time savings spent working on primary or current job | 32.09 | 28.91 | 0.00 | 100.0 | 29.15 | 26.77 | 0.00 | 100.0 | 37.24 | 32.13 | 0.00 | 100.0 |
| videocalls_ preCOVID _percent | Before the pandemic, what percentage of your normal working day did you spend in video calls? | 30.13 | 30.86 | 0.00 | 100.0 | 35.42 | 32.78 | 0.00 | 100.0 | 21.60 | 25.84 | 0.00 | 100.0 |
| videocalls_ current _percent | Currently, what percentage of your normal working day do you spend in video calls? | 39.04 | 30.82 | 0.00 | 100.0 | 42.18 | 31.52 | 0.00 | 100.0 | 33.52 | 28.47 | 0.00 | 100.0 |
| wfh_invest _burs | Percent of money invested in equipment or infrastructure enabling WFH that was paid for or reimbursed by Firm. Missing if no WFH or zero investment | 61.87 | 35.19 | 0.00 | 100.0 | 55.98 | 36.63 | 0.00 | 100.0 | 71.21 | 31.36 | 0.00 | 100.0 |
| wfh_hours invest | Hours invested in learning how to WFH effectively | 13.36 | 19.78 | 0.00 | 200.0 | 15.05 | 22.68 | 0.00 | 200.0 | 11.13 | 15.55 | 0.00 | 200.0 |
| educ_years | Years of education | 16.40 | 2.08 | 10.00 | 21.00 | 16.67 | 2.09 | 10.00 | 21.00 | 15.93 | 2.04 | 10.00 | 21.00 |
| age_quant | Age in years | 40.42 | 10.10 | 25.00 | 57.00 | 39.79 | 9.43 | 25.00 | 57.00 | 41.65 | 11.00 | 25.00 | 57.00 |
| workteam_ tasks_ percent | To perform your job, what percentage of your tasks require collaboration as part of a team? | 50.49 | 31.06 | 0.00 | 100.0 | 51.21 | 31.39 | 0.00 | 100.0 | 48.77 | 30.29 | 0.00 | 100.0 |
| hours_cc _partner | Currently, how many hours of childcare each week are provided by your partner? | 16.91 | 22.04 | 0.00 | 100.0 | 17.27 | 22.48 | 0.00 | 100.0 | 16.52 | 22.04 | 0.00 | 100.0 |
| hours_cc _other | Currently, how many hours of childcare each week are | 7.15 | 15.92 | 0.00 | 100.0 | 6.60 | 14.95 | 0.00 | 100.0 | 8.19 | 17.46 | 0.00 | 100.0 |

| | | | | | | | | | | | | |
|---|---|---|---|---|---|---|---|---|---|---|---|---|
| internet_ quality_ quant | provided by others, e.g. grandparents, babysitters? Internet quality - Fraction of time that internet works | 0.94 | 0.09 | 0.00 | 1.00 | 0.95 | 0.08 | 0.00 | 1.00 | 0.92 | 0.12 | 0.00 | 1.00 |
| numwfh _days_ Post COVID _boss_s_u | Firm planned share of paid working days WFH after COVID (%) | 48.12 | 41.78 | 0.00 | 100.0 | 49.79 | 41.30 | 0.00 | 100.0 | 44.78 | 42.95 | 0.00 | 100.0 |
| numwfh _days_ Post COVID _boss_pre | Firm planned share of paid working days WFH after COVID, before the most recent announcement made in the past 6 months (%) | 49.64 | 35.65 | 0.00 | 100.0 | 52.69 | 35.50 | 0.00 | 100.0 | 44.14 | 35.75 | 0.00 | 100.0 |
| coworker_ interactions | How much do you enjoy your personal interactions with coworkers at your Firm's worksite? | 7.62 | 2.60 | 0.00 | 10.00 | 7.80 | 2.57 | 0.00 | 10.00 | 7.38 | 2.64 | 0.00 | 10.00 |
| client_ interactions | How much do you enjoy your personal interactions with customers, clients, or patients at your Firm's worksite? | 8.04 | 2.50 | 0.00 | 11.00 | 8.25 | 2.35 | 0.00 | 11.00 | 7.72 | 2.60 | 0.00 | 11.00 |
| commute time_quant | Commute time (mins) | 37.70 | 32.44 | 0.00 | 120.0 | 41.93 | 35.04 | 0.00 | 120.0 | 31.50 | 27.55 | 0.00 | 120.0 |

*Categorical variables*

| | | Freq. | Percent | Cum. | Freq. | Percent | Cum. | Freq. | Percent | Cum. |
|---|---|---|---|---|---|---|---|---|---|---|
| employer _arr_qual | What plans does your Firm have for working arrangements of full-time employees after COVID, in 2022 or later? | | | | | | | | | |
| | 1 Fully on-site | 6,011 | 0.432 | 0.432 | | | | | | |
| | 2 Hybrid: 1 to 4 days WFH | 5,131 | 0.369 | 0.800 | | | | | | |
| | 3 Fully remote | 2,779 | 0.200 | 1.000 | | | | | | |
| gender | 1 Female | 5,634 | 0.405 | 0.405 | 1,918 | 0.319 | 0.319 | 1,487 | 0.535 | 0.535 |
| | 2 Male | 8,287 | 0.595 | 1.000 | 4,093 | 0.681 | 1.000 | 1,292 | 0.465 | 1.000 |
| | 3 Other/prefer not to say | 0 | 0.000 | 1.000 | 0 | 0.000 | 1.000 | 0 | 0.000 | 1.000 |
| race_ ethnicity | 1 Black or African American | 1,530 | 0.110 | 0.110 | 579 | 0.096 | 0.096 | 359 | 0.129 | 0.129 |
| | 2 Hispanic (of any race) | 1,180 | 0.085 | 0.195 | 454 | 0.076 | 0.172 | 260 | 0.094 | 0.223 |
| | 3 Asian | 805 | 0.058 | 0.252 | 291 | 0.048 | 0.220 | 191 | 0.069 | 0.291 |
| | 4 Native American or Alaska Native | 158 | 0.011 | 0.264 | 61 | 0.010 | 0.230 | 42 | 0.015 | 0.307 |
| | 5 Native Hawaiian or Pacific Islander | 45 | 0.003 | 0.267 | 22 | 0.004 | 0.234 | 12 | 0.004 | 0.311 |
| | 6 White (non-Hispanic) | 10,039 | 0.721 | 0.988 | 4,537 | 0.755 | 0.989 | 1,880 | 0.677 | 0.987 |
| | 7 Other | 164 | 0.012 | 1.000 | 67 | 0.011 | 1.000 | 35 | 0.013 | 1.000 |
| work_ industry | 1 Agriculture | 164 | 0.012 | 0.012 | 83 | 0.014 | 0.014 | 20 | 0.007 | 0.007 |
| | 2 Arts and Entertainment | 248 | 0.018 | 0.030 | 91 | 0.015 | 0.029 | 71 | 0.026 | 0.033 |
| | 3 Finance and Insurance | 2,260 | 0.162 | 0.192 | 1,016 | 0.169 | 0.198 | 401 | 0.144 | 0.177 |
| | 4 Construction | 801 | 0.058 | 0.249 | 388 | 0.065 | 0.263 | 125 | 0.045 | 0.222 |
| | 5 Education | 1,341 | 0.096 | 0.346 | 585 | 0.097 | 0.360 | 275 | 0.099 | 0.321 |
| | 6 Health Care and Social Assistance | 1,434 | 0.103 | 0.449 | 558 | 0.093 | 0.453 | 347 | 0.125 | 0.446 |
| | 7 Hospitality and Food Services | 277 | 0.020 | 0.469 | 113 | 0.019 | 0.471 | 59 | 0.021 | 0.467 |
| | 8 Information | 1,498 | 0.108 | 0.576 | 749 | 0.125 | 0.596 | 218 | 0.078 | 0.546 |
| | 9 Manufacturing | 880 | 0.063 | 0.640 | 367 | 0.061 | 0.657 | 175 | 0.063 | 0.608 |
| | 10 Mining | 128 | 0.009 | 0.649 | 64 | 0.011 | 0.668 | 21 | 0.008 | 0.616 |
| | 11 Professional and Business Services | 2,573 | 0.185 | 0.834 | 1,105 | 0.184 | 0.852 | 521 | 0.187 | 0.804 |
| | 12 Real Estate | 202 | 0.015 | 0.848 | 73 | 0.012 | 0.864 | 56 | 0.020 | 0.824 |
| | 13 Retail Trade | 554 | 0.040 | 0.888 | 217 | 0.036 | 0.900 | 128 | 0.046 | 0.870 |

| | | | | | | | | | |
|---|---|---|---|---|---|---|---|---|---|
| | 14 Transportation and Warehousing | 293 | 0.021 | 0.909 | 129 | 0.021 | 0.921 | 67 | 0.024 | 0.894 |
| | 15 Utilities | 199 | 0.014 | 0.923 | 75 | 0.012 | 0.934 | 50 | 0.018 | 0.912 |
| | 16 Wholesale Trade | 151 | 0.011 | 0.934 | 52 | 0.009 | 0.942 | 37 | 0.013 | 0.925 |
| | 17 Government | 737 | 0.053 | 0.987 | 273 | 0.045 | 0.988 | 167 | 0.060 | 0.985 |
| | 18 Other | 181 | 0.013 | 1.000 | 73 | 0.012 | 1.000 | 41 | 0.015 | 1.000 |
| wfh_ ownroom_ notbed | 100 Has their own room (not bedroom) to work in while WFH during COVID) | 6,848 | 0.492 | 0.492 | 2,392 | 0.398 | 0.398 | 1,784 | 0.642 | 0.642 |
| | 0 Otherwise | 7,073 | 0.508 | 1.000 | 3,619 | 0.602 | 1.000 | 995 | 0.358 | 1.000 |
| Live _children | Do you currently live with children under 18? -- categorical by youngest's age | | | | | | | | | |
| | 1 | 5,212 | 0.374 | 0.374 | 1,791 | 0.298 | 0.298 | 1,393 | 0.501 | 0.501 |
| | 2 | 3,547 | 0.255 | 0.629 | 1,652 | 0.275 | 0.573 | 589 | 0.212 | 0.713 |
| | 3 | 3,302 | 0.237 | 0.866 | 1,675 | 0.279 | 0.851 | 485 | 0.175 | 0.888 |
| | 4 | 1,036 | 0.074 | 0.941 | 506 | 0.084 | 0.936 | 164 | 0.059 | 0.947 |
| | 5 | 824 | 0.059 | 1.000 | 387 | 0.064 | 1.000 | 148 | 0.053 | 1.000 |
| disability_ qual | Do you have a health problem or a disability which prevents work or which limits the kind or amount of work you do? | | | | | | | | | |
| | 1 Yes | 2,872 | 0.206 | 0.206 | 1,603 | 0.267 | 0.267 | 332 | 0.119 | 0.119 |
| | 2 No | 10,857 | 0.780 | 0.986 | 4,359 | 0.725 | 0.992 | 2,394 | 0.861 | 0.968 |
| | 3 Prefer not to answer | 192 | 0.014 | 1.000 | 49 | 0.008 | 1.000 | 53 | 0.019 | 1.000 |

*Table A3 Regression results for Race & Industry*

| Dependent Variables | Stage 1 | | | Stage 2 | | | | | |
|---|---|---|---|---|---|---|---|---|---|
| | # Employees in the main work team | | | Predicted score: *Efficiency* in productivity | | | Predicted score: *Performance-revealed-by-promotion* | | |
| Variables | (1) All | (2) On-Site Favorable | (3) WFH-friendly | (1) All | (2) On-Site Favorable | (3) WFH-friendly | (1) All | (2) On-Site Favorable | (3) WFH-friendly |
| *Race* | | | | | | | | | |
| Hispanic (of any race) (reference: Black or African American) | -0.866 (0.631) | -0.696 (1.061) | -1.108 (1.252) | **-0.064*** **(0.033)** | **-0.097*** **(0.054)** | **0.057** **(0.068)** | -0.025 (0.033) | -0.064 (0.055) | 0.039 (0.066) |
| Asian | -0.017 (0.714) | 0.909 (1.221) | -1.122 (1.389) | **-0.181**** **(0.037)** | **-0.227**** **(0.062)** | **-0.078** **(0.075)** | 0.042 (0.037) | 0.094 (0.063) | 0.024 (0.074) |
| Native American or Alaska Native | 0.510 (1.367) | -0.914 (2.288) | 1.326 (2.522) | **0.029** **(0.071)** | **0.009** **(0.117)** | **0.151** **(0.137)** | 0.081 (0.071) | 0.117 (0.118) | 0.066 (0.134) |
| Native Hawaiian or Pacific Islander | 0.165 (2.466) | -2.329 (3.672) | -0.883 (4.519) | **0.131** **(0.129)** | **0.225** **(0.187)** | **0.170** **(0.245)** | -0.018 (0.129) | -0.057 (0.190) | 0.099 (0.240) |
| White (non-Hispanic) | 1.178** (0.461) | 1.246 (0.770) | 0.719 (0.905) | **-0.028** **(0.024)** | **-0.073*** **(0.039)** | **0.058** **(0.049)** | -0.033 (0.024) | -0.014 (0.040) | -0.061 (0.048) |
| Other | 1.451 (1.342) | 2.523 (2.189) | 2.598 (2.733) | **-0.155**** **(0.070)** | **-0.119** **(0.112)** | **-0.182** **(0.148)** | -0.139** (0.070) | -0.141 (0.113) | 0.056 (0.145) |
| *Industry* | | | | | | | | | |
| Arts and Entertainment (ref: Agriculture) | -3.458** (1.644) | -3.641 (2.572) | 0.219 (3.888) | -0.051 (0.086) | -0.058 (0.131) | 0.018 (0.211) | 0.120 (0.086) | 0.127 (0.133) | 0.057 (0.206) |
| Finance and Insurance | -0.938 (1.321) | -0.773 (1.939) | 3.384 (3.520) | 0.036 (0.069) | -0.035 (0.099) | 0.021 (0.191) | 0.206*** (0.069) | 0.202** (0.100) | 0.145 (0.187) |
| Construction | 0.200 (1.399) | -0.275 (2.050) | 4.937 (3.699) | -0.094 (0.073) | -0.180* (0.105) | -0.026 (0.201) | 0.170** (0.073) | 0.113 (0.106) | 0.199 (0.196) |
| Education | -2.102 (1.358) | -2.752 (2.003) | 3.151 (3.572) | -0.228*** (0.071) | -0.345*** (0.102) | -0.091 (0.194) | 0.163** (0.071) | 0.125 (0.104) | 0.137 (0.190) |
| Health Care and Social Assistance | -0.978 (1.352) | -0.334 (2.006) | 3.450 (3.541) | -0.044 (0.071) | -0.132 (0.102) | 0.096 (0.192) | 0.167** (0.070) | 0.122 (0.104) | 0.091 (0.188) |
| Hospitality and Food Services | -1.977 (1.610) | -2.592 (2.455) | 4.212 (3.981) | -0.222*** (0.084) | -0.356*** (0.125) | 0.069 (0.216) | 0.221*** (0.084) | 0.343*** (0.127) | 0.083 (0.211) |
| Information | 0.587 (1.345) | 1.821 (1.967) | 4.206 (3.594) | 0.070 (0.070) | 0.005 (0.100) | 0.158 (0.195) | 0.172** (0.070) | 0.170* (0.102) | 0.010 (0.191) |
| Manufacturing | 1.108 (1.389) | 1.641 (2.062) | 5.169 (3.627) | -0.090 (0.073) | -0.141 (0.105) | 0.050 (0.197) | 0.113 (0.072) | 0.103 (0.107) | 0.034 (0.193) |
| Mining | 4.966*** (1.923) | 4.264 (2.816) | 4.271 (4.795) | -0.128 (0.101) | -0.286** (0.144) | 0.036 (0.260) | 0.058 (0.100) | 0.117 (0.146) | -0.318 (0.255) |
| Professional and Firm Services | 1.536 (1.319) | 1.353 (1.939) | 5.613 (3.509) | -0.056 (0.069) | -0.137 (0.099) | 0.008 (0.190) | 0.199*** (0.069) | 0.201** (0.100) | 0.121 (0.186) |
| Real Estate | -0.235 (1.716) | 0.155 (2.719) | 2.689 (4.008) | -0.018 (0.090) | 0.028 (0.139) | 0.138 (0.217) | 0.213** (0.089) | 0.223 (0.141) | 0.087 (0.213) |
| Retail Trade | 0.428 (1.452) | -0.245 (2.189) | 4.769 (3.695) | -0.166** (7.589) | -0.299*** (0.112) | -0.049 (0.200) | 0.094 (0.076) | 0.121 (0.113) | 0.020 (0.196) |
| Transportation and Warehousing | 3.539** (1.591) | 5.366** (2.380) | 8.868** (3.917) | -0.199** (0.083) | -0.371*** (0.121) | 0.175 (0.212) | 0.236*** (0.083) | 0.280** (0.123) | 0.056 (0.208) |
| Utilities | 1.712 (1.719) | 1.086 (2.694) | 8.693** (4.058) | -0.045 (0.090) | -0.006 (0.137) | 0.051 (0.220) | 0.294*** (0.090) | 0.309** (0.139) | 0.183 (0.215) |
| Wholesale Trade | 0.931 (1.841) | -0.661 (2.994) | 9.547** (4.273) | -0.143 (0.096) | -0.198 (0.153) | -0.113 (0.232) | 0.227** (0.096) | 0.152 (0.155) | 0.372 (0.227) |
| Government | 2.505* (1.417) | 2.749 (2.141) | 4.396 (3.644) | -0.124* (0.074) | -0.211* (0.109) | -0.045 (0.198) | 0.206*** (0.074) | 0.269** (0.111) | 0.051 (0.193) |
| Other | 2.153 (1.761) | 3.512 (2.724) | 4.772 (4.194) | -0.038 (0.092) | -0.160 (0.139) | -0.197 (0.227) | 0.132 (0.092) | 0.001 (0.141) | 0.160 (0.223) |

## Appendix reference